\documentclass[12pt,draftclsnofoot,onecolumn]{IEEEtran}
\usepackage{graphicx}
\usepackage{amsmath}
\usepackage{amssymb}
\usepackage[caption=false]{subfig}
\usepackage[noadjust]{cite}
\usepackage{float}
\usepackage{color}
\usepackage{url}
\usepackage{graphicx}
\usepackage{amsfonts}
\usepackage{float}
\usepackage{color}
\usepackage{cite}
\usepackage{epstopdf}
\usepackage{bm}
\usepackage{mathtools}
\usepackage[algoruled,resetcount,linesnumbered]{algorithm2e}
\usepackage{array}
\usepackage{multirow}
\usepackage{multicol}
\usepackage{mathtools}
\newtheorem{remark}{Remark}

\graphicspath{ {fig/} } %This is where you have saved the images

\begin{document}

\title{Off-Grid Aware Channel and Covariance Estimation in mmWave Networks}
% \title{Off-Grid Aware mmWave Communications}

\author{Chethan Kumar Anjinappa,
Ali Cafer Gurbuz, Yavuz Yap{\i}c{\i}, and \.{I}smail G\"{u}ven\c{c} 
\thanks{C.K. Anjinappa, Y. Yapici, and \.{I}. G\"{u}ven\c{c} are with the Department of Electrical and Computer Engineering, North Carolina State University, Raleigh, NC (e-mail:~\{canjina,yyapici,iguvenc\}@ncsu.edu).}
\thanks{A.C. G\"{u}rb\"{u}z is with the the Department of Electrical and Computer Engineering, Mississippi State University, Mississippi State, MS (e-mail:~gurbuz@ece.msstate.edu).}}

\maketitle

\begin{abstract}
%\textcolor{red}{New abstract, thanks to Dr. Yavuz}
The spectrum scarcity at sub-6 GHz spectrum has made millimeter-wave (mmWave) frequency band a key component of the next-generation wireless networks. While mmWave spectrum offers extremely large transmission bandwidths to accommodate ever-increasing data rates, unique characteristics of this new spectrum need special consideration to achieve the promised network throughput. In this work, we consider the \textit{off-grid} problem for mmWave communications, which has a significant impact on basic network functionalities involving beam steering and tracking. The off-grid effect naturally appears in compressed sensing (CS) techniques adopting a discretization approach for representing the angular domain. This approach yields a finite set of discrete angle points, which are an \textit{approximation} to the \textit{continuous} angular space, and hence degrade the accuracy of related parameter estimation. In order to cope with the off-grid effect, we present a novel parameter-perturbation framework to efficiently estimate the channel and the covariance for mmWave networks. The proposed algorithms employ a smart perturbation mechanism in conjunction with a low-complexity greedy framework of simultaneous orthogonal matching pursuit (SOMP), and jointly solve for the off-grid parameters and weights. Numerical results show a significant performance improvement through our novel framework as a result of handling the off-grid effects, which is totally ignored in the conventional sparse mmWave channel or covariance estimation algorithms.
%This work investigates two critical problems in  millimeter (mmWave) hybrid multiple-input multiple-output (MIMO) networks: \textit{channel estimation} and \textit{spatial covariance estimation} with an emphasis on the \textit{off-grid} effect. The off-grid effect is prevalent in the compressed sensing (CS) schemes which adopt  a discretization procedure. In such an approach, the discretization procedure yields a finite discrete point which is an approximation to the \textit{continuous} parametric space degrading the estimation accuracy. The accuracy of these estimation plays a critical role in applications like channel tracking, beam steering, etc. To eliminate this off-grid effects, we present a novel parameter-perturbation framework for both the channel and covariance estimation problems. The designed algorithms are used for the \textit{spatial channel covariance matrix estimation} which is the central goal of this work. The proposed algorithm uses an efficient controlled perturbation mechanism in conjunction with the popular low-complexity greedy framework simultaneous orthogonal matching pursuit (SOMP), to jointly solve for the off-grid parameters and weights. Numerical results verify the performance improvement through our proposed algorithms both in terms of the \textit{relative efficiency} and \textit{reconstruction error} metrics, which is basically due to solving for the off-grid effects carefully, which is ignored in the conventional CS reconstruction algorithms.
\end{abstract}
%The designed algorithms are evaluated based on the relative energy efficiency metric and the reconstruction error.The efficacy of our proposed algorithms are demonstrated by the simulations.


\section{Introduction}\label{sec:intro}
One of the most promising aspects of next generation wireless networks is the use of high-frequency large-bandwidth signals in the millimeter-wave (mmWave) frequency bands. The small wavelengths at these frequencies enable the use of a large number of antennas (dozens to hundreds) within a small physical area. %reasonable physical form factor, 
This helps to compensate for the high path losses, making the multiple-input multiple-output (MIMO) a quintessential technology at mmWave bands \cite{roh2014millimeter,rangan2014millimeter}. As a consequence, MIMO technology has a great potential in mmWave networks to deliver higher data rates, higher spectral efficiency, and lower latency, highly exceeding what is possible with the traditional cellular networks operating at sub-6 GHz bands. 

In the conventional MIMO architecture, use of large number of antennas results in high cost and high power consumption which makes it difficult to assign a distinct radio-frequency (RF) chain per antenna. To curtail these issues, \textit{hybrid analog/digital beamforming (HADB)} architecture is typically adapted at the mmWave bands \cite{mendez2016hybrid,heath2016overview}. In \textit{HADB} architecture, the MIMO processing is split between the analog RF and digital baseband (BB) domains to reduce the number of required transceivers. One of the critical challenges with this architecture is to effectively configure the analog precoding stages. 

In most of the prior work, the problem of configuring the analog precoder is accomplished assuming the availability of  full channel state information (CSI) at the transmitter which is difficult to achieve even for a time division duplexing network. As a promising alternative to full CSI, a \textit{spatial covariance matrix} based method has recently been proposed to update the analog RF precoders \cite{park2016spatial,mendez2015adaptive,park2017spatial}. Further, the low dimensional measurements due to the limited RF chains makes it difficult to obtain the accurate \textit{channel} and \textit{spatial covariance} estimates, which has a significant impact on the basic network functionalities involving beam alignment \cite{Beam_Alignment_Song_Caire} and tracking \cite{MIMO_Channel_Tracking}. To overcome these challenges, algorithms based on compressed sensing (CS) \cite{Uniform_Sampling_Cos_Domain,2014_Heath_Channel_Estimation_mmWave,MIMO_Channel_Tracking,Beam_Alignment_Song_Caire} have been proposed. However, these algorithms ignore the off-grid effects which is prevalent in the CS schemes \cite{tang2013compressed_off_grid,chi2011sensitivity}. %In this work, we consider the off-grid problem for mmWave communications, which has a significant impact on basic network functionalities involving beam steering and tracking.

%Furthermore, the low dimensional measurements due to the limited RF chains makes it even more difficult to estimate the channel accurately. To overcome these challenges, channel estimation algorithms based on compressed sensing (CS) \cite{Uniform_Sampling_Cos_Domain,2014_Heath_Channel_Estimation_mmWave} have been proposed.  However, these methods are only suited for time-invariant/non-frequency selective channels.

%\subsection{Our Contributions}
In this work, we consider the off-grid problem for mmWave communications and propose two novel parameter perturbed algorithms for the off grid channel and spatial covariance estimation problems, respectively. Specifically, our contributions in this paper are the following:
\begin{itemize}
    \item \textit{Off-Grid Aware Channel Estimation:} Motivated by the spirit behind \cite{2018_POMP_ICC} which focuses on the single measurement vector (SMV) setup, we extend the parameter perturbed orthogonal matching pursuit (PPOMP) based channel estimation to the multiple measurement vector (MMV) setup. The peculiarity of this work is the MMV case which is aided by the simultaneous OMP (SOMP) framework, and the inclusion of the non-apparent non-uniform sampling of the physical domain discussed in Section~\ref{SubSub:Discretization}. The SMV framework can be considered as a special case of the MMV framework presented in this work. 
\item \textit{Off-Grid Aware Covariance Estimation:}
To the best of authors' knowledge this paper presents the first off-grid aware explicit covariance estimation method for mmWave MIMO networks. More specifically, the algorithm is designed for both the uniform/non-uniform sampling schemes employed in the discretization procedure and exploits the inherent Hermitian property of the covariance matrix.
\end{itemize}

The proposed algorithms evade the issue arising from the basis mismatch problems by operating on the continuum angle-of-arrival (AoA) and angle-of-departure (AoD) space using the mechanism of the controlled perturbation in conjunction with a modified SOMP framework. The SOMP framework helps to preserve the low computational complexity which is inherent for a greedy solver. The key in the designed parameter-perturbed framework is to preserve the sub-optimal greedy projection step of the SOMP algorithm and then invoke controlled perturbation mechanism on the selected columns from the projection step. This procedure allows one to combat the off-grid effects after the projection step and before the update of the residual terms which is the central innovation behind both the developed parameter perturbed algorithms. We present the rationale behind this central innovation and validate the superiority of the proposed methods by numerical simulations.

The remainder of the paper is organized as follows. Section \ref{Sec:Literature_Review} presents a brief literature review on spatial covariance based hybrid precoding and off-grid effects in the CS schemes. Section \ref{sec:system} presents the time-varying system/channel model followed by the uniform/non-uniform sampling schemes for the discretization procedure and problem formulation in Section \ref{Sec:Sparse_Representation}. In Section \ref{Sec: Proposed Soltuion}, we present the parameter perturbed framework for the channel estimation problem which uses controlled perturbation mechanism in conjunction with the SOMP framework. We then extend this framework to the covariance estimation problem in Section \ref{Sec:Explicit_Covariance_Estimation}. In Section \ref{sec:results}, we validate the efficacy of our proposed algorithms using computer simulations, and finally, we provide concluding remarks in Section~\ref{sec:conclusion}.

{\bf Notation}: Vectors and matrices are represented by lower-case (eg: \textbf{a}) and upper-case boldface (eg: \textbf{A}) letters, respectively. Every vector is considered as a column vector. The transpose, conjugate, conjugate transpose, and pseudo-inverse of a matrix \textbf{A} are denoted by  $\textbf{A}^\text{T}$, $\textbf{A}^\text{H}$, $\textbf{A}^{*}$, and $\textbf{A}^{\dagger}$, respectively. $\mathop{\mathbb{E}}(\cdot)$ is the expectation operator. For an integer $K$, we use the shorthand notation $[K]$ for the set of non-negative integers $\{1,2,\ldots, K\}$. The support of a vector $\textbf{x} \in \mathcal{C}^N$ is the index set of non-zero entries of $\textbf{x}$, i.e., supp(\textbf{x}) = $\{j \in [N]: x_j \neq 0\}$. The vector \textbf{x} is called $k$-sparse if at most $k$ of its entries are non-zero. For $M \times N$ matrix \textbf{A} and $P \times Q$ matrix \textbf{B}, \textbf{A} $\bigotimes$ \textbf{B} denotes the $M P \times N Q$ matrix of Kronecker product. \textbf{A} $\overset{(K)}{\bigodot}$ \textbf{B} = [$\textbf{A}_1 \bigotimes \textbf{B}_1 \ldots \textbf{A}_K \bigotimes \textbf{B}_K$] denotes the generalized Khatri-Rao product with respect to $K$ partitions where \textbf{A} = [$\textbf{A}_1 \ldots \textbf{A}_K$] and \textbf{B} = [$\textbf{B}_1 \ldots \textbf{B}_K$]. We use $\sim$ notation to denote ``is distributed as". Finally, the $\mathbb{I}_N$ denote the identity matrix of size $N \times N$.
%$\boldsymbol{\Gamma} = diag(\boldsymbol{\gamma})$ represents the diagonal matrix which maps an N-tuple $\boldsymbol{\gamma}$ = $[\gamma_1,\gamma_2,\dots,\gamma_N]$ to the corresponding diagonal matrix.

\section{Literature Review}\label{Sec:Literature_Review}
\subsection{Spatial Covariance based Hybrid Precoding and Related Work}
The spatial covariance exploits the relatively stationary long-term statistics of the propagation channel, and it can be leveraged for precoder design in mmWave networks~\cite{li2017optimizing,mendez2015adaptive,JSDM_Caire,park2016spatial,park2017spatial}. The rationale behind the use of spatial covariance matrix are two-fold. Firstly, in many cases, the angular coherence time (several seconds or more) is much longer than the channel coherence time (several milliseconds) \cite{va2017impact,li2017optimizing}. As a result, the angular and average power features of the channel can be assumed to be time-invariant, resulting in the spatial covariance matrix to be constant across many channel coherence intervals. Secondly, the spatial covariance matrix is frequency invariant, due to the significant angular congruence across the frequency bands \cite{Angular_Temporal_Correlation_Chethan_2018,Out_of_band}, which is important for a wideband system where a common analog precoder can be shared across different sub-carriers. These reasons make the spatial covariance based precoding particularly attractive: once the RF beamformer is designed based on the channel covariance, it need not be updated every time instant. We would like to refer the reader to the works \cite{park2016spatial,mendez2015adaptive,li2017optimizing,park2017spatial} for a comprehensive discussion on the spatial covariance estimation for mmWave HADB MIMO architectures. 

%However, estimating the covariance is complicated due to the fact that only the signals pre-combined by the analog precombiner are available at the baseband. Next subsection describes the relevant works in spatial covariance matrix estimation in mmWave networks.

Estimating the covariance is complicated due to the fact that only the signals pre-combined by the analog precombiner are available at the baseband.
Based on the way the covariance matrix is estimated, it can be broadly categorized into two methods: 1) \textit{covariance estimation via the channel estimation framework} which we will refer as the indirect method, and 2)  \textit{explicit covariance estimation} which we will refer as the direct method hereafter. The central idea in the indirect approach is to solve for the channel estimates for every successive snapshot and use these estimates to calculate the covariance matrix. Upon obtaining the channel estimates for every snapshot, the covariance calculation is relatively straightforward. However, in cases when the channel estimates are not required, then one can explicitly operate on the covariance of measurements directly to estimate the covariance matrix which is central to the latter approach. Both the \textit{channel estimation} and the \textit{covariance estimation} problems can be posed as a compressed sensing (CS) problem leveraging the sparse nature of mmWave channels~\cite{ChannelModel_Rappaport1,Vector_Type,2018_POMP_ICC,park2018spatial,Out_of_band}.

In the literature, several CS approaches have been utilized to estimate the channel and the spatial covariance. For the indirect approach, the channel estimates can be obtained using the SMV CS techniques such as \cite{Vector_Type,2018_POMP_ICC}. However, these SMV techniques fail to exploit the common support of the channel estimates across different snapshots. The common support across multiple snapshots is due to the invariant angular domain features across multiple snapshots which is central to the use of spatial covariance matrix. The MMV techniques can exploit this common support structure; however, most of the MMV techniques are designed with sensing matrix fixed over all the snapshots making it inefficient for \textit{time-varying sensing} matrices. The statistical problem of covariance estimation can be approached by explicitly estimating the covariance using the measurement covariance space. Strategies such as MUSIC \cite{MUSIC} and ESPRIT \cite{ESPIRIT} algorithms can be adopted but these methods fail to leverage the channel sparsity. Recently, a CS MMV based covariance estimation for the time-varying sensing matrices has been proposed in \cite{park2018spatial} and a tensor-based decomposition approach has been proposed in \cite{2018_Tensor}. Further CS algorithms for the direct approach of spatial covariance estimation can be found in \cite{park2017spatial,park2018spatial,Out_of_band,Relative_Metric_Eta_Caire,2018_Tensor}

\subsection{Off-Grid Effects and Related Work}
The CS-based methods discussed in Section~\ref{Sec:Literature_Review} are based on the concept of \textit{virtual channel models} \cite{sayeed2002deconstructing}, which provide a virtual angular representation of MIMO channels employing a discretization procedure.
The discretization procedure results in an exact sparse representation of the virtual channel model only when the true AoA and AoD lies on one of the pre-defined set of spatial angles employed during the discretization. However, the true AoA-AoD lies in the continuous space and may not fall exactly onto one of the finite pre-defined spatial angles. In fact, for the discrete Fourier transform (DFT) basis defined by the virtual channel model, a continuous AoA-AoD parameter lying between two successive DFT grid cells will affect not the only the closest two cells, but the whole grid with amplitude decaying with $1/{N_\text{AoA}N_\text{AoD}}$ due to the \textit{Dirichlet kernel} \cite{teke2013perturbed,Dirichlet_Kernel}, where $N_\text{AoA}$ and $N_\text{AoD}$ are the number of grid points in the AoA and AoD grid, respectively. This \textit{off-grid} phenomena violates the sparsity assumption, resulting in a decrease in reconstruction performance. As a result, the estimation accuracy of the CS based methods is limited by the number of grid points \cite{tang2013compressed_off_grid,chi2011sensitivity,teke2013perturbed,Dirichlet_Kernel}. 

A natural approach to the problem of off-grid/basis mismatch is to increase the number of grid points corresponding to decrease in grid sizes. However, this is an inefficient approach due to the following two main problems: Firstly, it increases the mutual coherence of the dictionary, violating the restricted isometric property \cite{candes2008restricted}, which makes it more difficult to reconstruct using standard compressed sensing analyses. Further, it also increases the dimension of the dictionary and the sparse vector to be recovered, resulting in higher memory and computational complexity in reconstruction. More details on the basis mismatch/off grid effects can be found in the seminal paper \cite{tang2013compressed_off_grid} and further discussion in \cite{chi2011sensitivity,teke2013perturbed,teke2014robust} with a focus on applications such as beamforming, radars, and image reconstruction. 

An alternative is to tackle the off-grid effects upfront without increasing the grid size. For example, in the context of channel estimation, Tang et al. \cite{tang2019off} provide improved off-grid sparse Bayesian algorithm for the channel estimation framework. A grid-less CS technique is developed via atomic norm minimization in the form of semi-definite programming by Wang et al. \cite{2017_Gridless_Atomic}. Although these problems tackle the off-grid issues, the computational complexity of these methods are significantly high. In previous work, Gurbuz et al. provide a controlled perturbation mechanism for spatial angular parameters based on orthogonal matching pursuit (OMP) \cite{2018_POMP_ICC} but is tailored only for the SMV setup with the immediate application to the MMV setup being not straightforward. Also, the application of these off-grid methods to the covariance estimation problem is not straight forward. More importantly, to the best of our knowledge, there is no work which  investigates the off-grid effects or provide an off-grid based solution explicitly for the  covariance estimation problem. This motivates the development and analysis of robust low-complexity channel and covariance estimation techniques for the MMV setup with emphasis on basis mismatch effects.

%%%%%%%%%%%%%%%%%%%%%%%%%%%%%%%%%%%%%%%%%%%%%%%%%%%%%%%%%%%%%%%%%%%%%%%%%%%%%%%%%%%%%%%%%%%%%%%%%%%%%%%%%%%%%%%%%%%%%

\section{System and Channel Model}\label{sec:system}
%In this section, we present the time-varying system and channel model followed by the problem formulations of both channel and spatial covariance estimations.
\subsection{System Model}
\begin{figure}[!t]
    \centering
    \includegraphics[scale=0.4]{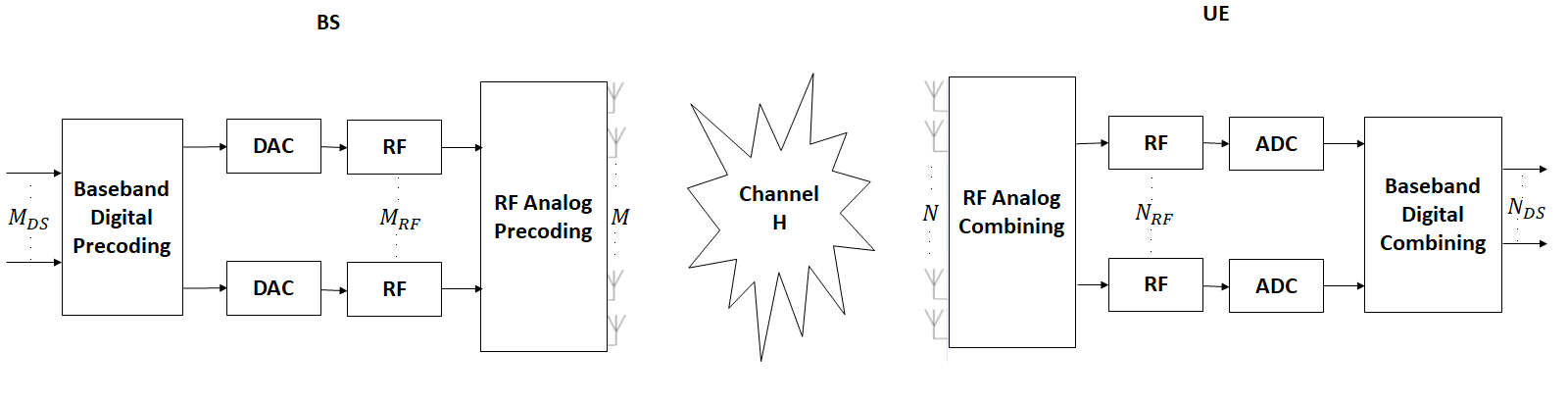}
    \caption{Block diagram of a mmWave HADB MIMO system.}
    \label{fig:MIMO}
\end{figure}

Consider a HADB mmWave MIMO network comprised of a base station (BS) communicating with a generic user equipment (UE), both equipped with a uniform linear array (ULA). We assume the BS is equipped with $M$ antennas, $M_\text{RF}$ RF chains, and  $M_\text{DS}$ data streams. Similarly, the UE is assumed to be equipped with $N$ antennas, $N_\text{RF}$ RF chains, and $N_\text{DS}$ data streams to guarantee multi-stream data transmission. Typically, it is assumed that $M_\text{DS} \leq M_\text{RF}\leq M$ and $N_\text{DS} \leq N_\text{RF}\leq N$. This is visualized in Fig. \ref{fig:MIMO}. For the system and channel model, we follow the model adopted by \cite{park2018spatial} which we refer to as ``\textit{time-varying sensing matrix for the time varying channel}" model. The \textit{time-varying sensing matrix} model is detailed up next, while the \textit{time-varying channel model} will be discussed further in Section \ref{SubSec:Channel_Model}.

During the training period, at frame $t$ each symbol $s$ is transmitted in individual succession through its dedicated RF chain resulting in a total of $M_\text{RF}$ consecutive training symbols from the BS per frame. During these $M_\text{RF}$ successive symbols transmission the channel $\mathbf{H}_t$ is assumed to be invariant, implying the channel $\mathbf{H}_{t,s} = \mathbf{H}_{t},\forall s\in [M_\text{RF}] $ will be constant across all the symbols $s$ for frame $t$. However, the channel $\mathbf{H}_t$ can change across the frames. More details on the channel model is presented in Section \ref{SubSec:Channel_Model}. For every symbol $s$ at frame $t$, the transmitter uses pilot beam patterns through one of its dedicated RF chain denoted by the precoder operation as $\{\mathbf{f}_{t,s} \in \mathcal{C}^{M\times 1}; ||\mathbf{f}_{t,s}||_2^2 = 1, \forall t \in [T], \forall s\in[M_\text{RF}]\}$. The receiver can use $N_\text{RF}$ beam pattern simultaneously denoted by the combiner operation denoted as $\{\mathbf{W}_{t,s}= [\mathbf{w}_{t,s,1},\ldots, \mathbf{w}_{t,s,N_\text{RF}}] \in \mathcal{C}^{N \times N_\text{RF}}; ||\mathbf{w}_{t,s,i}||_2 = 1,  \forall t \in [T], \forall s\in[M_\text{RF}], \forall i \in [N_\text{RF}]\}$. With this notation, the received signal at the baseband during the symbol $s$ at frame $t$ is given by
\begin{equation}\label{y_Equ}
\mathbf{y}_{t,s} = \mathbf{W}_{t,s}^\text{H} \mathbf{H}_t \mathbf{f}_{t,s} \mathbf{x} + \mathbf{W}_{t,s}^\text{H} \mathbf{n}_{t,s}, \quad t \in [T], s \in [M_\text{RF}],
\end{equation}
where $\mathbf{H}_t \in \mathcal{C}^{N \times M}$ represents the time-varying channel matrix from the BS to UE. $\mathbf{n}_{t,s} \in \mathcal{C}^{N \times 1}$ denotes the noise vector at the UE which is assumed to be a circularly symmetric \textit{i.i.d} Gaussian distributed, $\mathbf{n}_{t,s} \sim \mathcal{CN}(0,\sigma_n^2\mathbb{I}_N)$, where $\sigma_n^2$ is the noise variance. The transmitted pilot symbol $\mathbf{x}$ is known at the BS, thereby omitted here onwards. The precoder and combiner operation are denoted by $\mathbf{f}_{t,s}$ and $\mathbf{W}_{t,s}$, respectively, which can be chosen to be either \textit{static} or \textit{dynamic} across different symbols $s$ for each frame $t$. 

For the static case, the precoder is given as $\mathbf{f}_{t,s} = \mathbf{f}_t,\forall s\in [M_\text{RF}]$, which implies that the same precoder $\mathbf{f}_t$ is used for all the symbols $s\in [M_\text{RF}]$ throughout the frame $t$. On the other hand, for the dynamic case, however, the precoder changes for every symbol $s$ in the frame $t$. Similarly, this is applicable for the combiner operation resulting in a total of four different combinations for the choice of precoder and combiner matrices. The work \cite{park2018spatial} established that the use of time-varying analog precoding/combining matrix across the symbols at each frame provides larger gain over fixed precoding/combining methods and increases the recovery success probability. We would like to refer the reader to the work \cite{park2018spatial} for a comprehensive discussion on the four different possibilities. Thus, throughout this work we assume both the precoder and combiner is time-varying for all the symbols at each frame and we restrict our discussion only to the \textit{dynamic} case hereafter.

With the above setting, the received signal $\mathbf{y}_{t,s}\in \mathcal{C}^{N_\text{RF}\times 1}$ in (\ref{y_Equ}) can be stacked together in rows, $[\mathbf{y}_{t,1}, \ldots , \mathbf{y}_{t,s}], \forall s \in [M_\text{RF}],  t \in [T]$, which we denote as ${\tilde{\mathbf{y}}}_{t,\text{agg}}$. The row-wise stack yields a $M_\text{RF} N_\text{RF} \times 1$ vector per frame and is mathematically represented as
\begin{align}\label{y_Equ_Agg}
{\tilde{\mathbf{y}}}_{t,\text{agg}} = { \left({\tilde{\mathbf{F}}}_{t,\text{agg}} \overset{(M_\text{RF})}{\bigodot} {\tilde{\mathbf{W}}}_{t,\text{agg}}^\text{T}\right)^\text{T} \text{vec} (\mathbf{H}_t)} + \mathbf{W}_{t,\text{agg}}^\text{H} \mathbf{n}_{t,\text{agg}}, \quad  t \in [T],
\end{align}
where $\tilde{\mathbf{W}}_{t,\text{agg}} = [\mathbf{W}_{1,1}^T, \mathbf{W}_{1,2}^T \ldots, \mathbf{W}_{t,s}^T]^T$, and ${\tilde{\mathbf{F}}}_{t,\text{agg}} = [\mathbf{f}_{t,1}, \ldots, \mathbf{f}_{t,s}];\forall \mathbf{s}\in [M_\text{RF}]$ are the aggregated version of the combiner, and precoder respectively. The $\overset{(M_\text{RF})}{\bigodot}$ denotes the generalized Khatri-Rao product with respect to $M_\text{RF}$ partitions, while $\mathbf{h}_t = \text{vec}(\mathbf{H}_t) \in \mathcal{C}^{NM \times 1}$ is the vectorized form of the channel matrix $\mathbf{H}_t$. Hereafter, $\mathbf{\Phi}_{t,\text{agg}} = \left(\mathbf{\tilde{F}}_{t,\text{agg}} \overset{(M_\text{RF})}{\bigodot} \mathbf{\tilde{W}}_{t,\text{agg}}^\text{T}\right)^\text{T}$ and $ \mathbf{\tilde{n}}_{t,\text{agg}} = \mathbf{W}_{t,\text{agg}}^\text{H} \mathbf{n}_{t,\text{agg}}$, then the resulting signal in (\ref{y_Equ_Agg}) can be rewritten as 
\begin{align}\label{y_Equ_Agg1}
\mathbf{\tilde{y}}_{t,\text{agg}} = \mathbf{\Phi}_{t,\text{agg}} \mathbf{h}_t + \mathbf{\tilde{n}}_{t,\text{agg}},  \quad \forall t \in [T].
\end{align}
\subsection{Channel Model}\label{SubSec:Channel_Model}
The mmWave channels can be well approximated by the geometric channel models \cite{ ChannelModel_Rappaport1,ChannelModel_Andreas_Molisch,Propogation_Statistics_Chethan_2018} which captures the natural spatial channel sparsity. In \cite{ChannelModel_Andreas_Molisch,Propogation_Statistics_Chethan_2018} it is shown that, even in highly non-line-of-sight (NLOS) environments, the communication between the BS and a UE potentially happens with multiple spatial clusters. Following the model in \cite{ChannelModel_Rappaport1}, we assume the channel to be composed of $K$ spatial path clusters with each cluster containing $L$ macro-level scattering multi-path components (MPCs) \cite{ChannelModel_Rappaport}. Note that $K$ and $L$ may each be time-varying due to mobility of the UE and the surrounding scatterers \cite{Angular_Temporal_Correlation_Chethan_2018,Propogation_Statistics_Chethan_2018}. However, for simplicity, we assume $K$ and $L$ to be fixed at least for the duration of the covariance estimation. Here on, we use the short notation of $K_L$ to represent a total of $KL$ MPCs.

Further, at mmWave bands, the coherence time of time-varying fading coefficients is much shorter than that of angular coherence time (the time scale over which the angular profile changes significantly) implying the significant time-variations of the channel coefficients even in moderate mobility \cite{Beamwidth_Temporal_Channel_Variataion_Heath}. Typically, the angular coherence time takes several seconds or more to change significantly relative to the coherence time which is on the order of several milliseconds. As a result, the spatial features of the channel can be assumed to be time-invariant or locally constant (or very slowly time-varying) and small-scale fading coefficients (complex path gains) are assumed to be varying much faster. This model is widely used in the literature and confirmed by several channel measurements and sounders  \cite{park2018spatial,Beam_Alignment_Song_Caire,Beamwidth_Temporal_Channel_Variataion_Heath}. Under the stated assumptions, the double directional time-varying channel matrix $\mathbf{H}_t$ at time frame $t$ can be expressed as
\begin{equation}\label{H_Equation}
\mathbf{H}_t = \frac{1}{\beta}\sum_{k=1}^K \sum_{l=1}^L \alpha_{k,l,t} \mathbf{a}_\text{UE}(\theta_{k,l}^\text{rx}) \mathbf{a}_\text{BS}(\theta_{k,l}^\text{tx})^\text{H},  %= \frac{1}{\beta} \bf{A}_R \bf{\Sigma} \bf{G}_t \bf{A}_T^H
\end{equation}
where $K$ denotes the number of clusters/scatterers, $L$ denotes the number of MPCs from each cluster, $\beta$ is the average path loss, $\alpha_{k,l,t}$ denotes the small scale fading time-varying complex gain of the $l^\text{th}$ MPC in the $k^\text{th}$ cluster during time frame $t \, \in [T]$ where $T$ is the total number of time frames (snapshots), while $\theta_{k,l}^\text{rx} \in [0,\pi)$ and $\theta_{k,l}^\text{tx}\in [0,\pi)$ denote the azimuthal AoA and AoD of the $l^\text{th}$ MPC in the $k^\text{th}$ cluster, respectively. Unlike $\alpha_{k,l,t}$, the AoA $\theta_{k,l}^\text{rx}$ and the AoD $\theta_{k,l}^\text{tx}$ for all the MPCs are assumed to be constant across the $T$ snapshots. 

The complex gain $\alpha_{k,l,t}$ are modeled as the $i.i.d$ random variable with the complex Gaussian distribution, $\alpha_{k,l,t} \sim \mathcal{CN}(0,1)$. Further, the AoA is expressed as $\theta_{k,l}^\text{rx} = \theta_{k}^\text{rx} + \zeta_{l}$, where $\theta_k^\text{rx}$ is distributed uniformly over $[0,\pi)$ and $\zeta_{l}$ follows a Laplacian distribution $\mathcal{L}(0,\sigma_\text{AS}^\text{AoA})$ with the zero mean and scaling parameter of $\sigma_\text{AS}^\text{AoA}$, where $\sigma_{AS}^\text{AoA}$ is the AoA angular spread. Likewise, $\sigma_{AS}^\text{AoD}$ is the AoD angular spread. 

The terms $\mathbf{a}_\text{BS}(\theta^\text{tx})$ and $\mathbf{a}_\text{UE}(\theta^\text{rx})$ in (\ref{H_Equation}) are the normalized array response to an MPC coming from the angles $\theta^\text{tx}$ and $\theta^\text{rx}$ with respect to (w.r.t) the BS and UE ULA, respectively. The normalized ULA responses at the BS and UE are expressed as 
\begin{eqnarray}\label{Eq:Array_Response}
\begin{aligned}
&[\mathbf{a}_\text{BS} (\theta^\text{tx})]_m=\frac{1}{\sqrt{M}}e^{j \frac{2 \pi }{\lambda}d_\text{BS}(m-1) \cos(\theta^\text{tx})}, \quad \forall m \in [M],\\
&[\mathbf{a}_\text{UE}(\theta^\text{rx})]_n = \frac{1}{\sqrt{N}} e^{j \frac{2 \pi}{\lambda}d_\text{UE}(n-1) \cos(\theta^\text{rx})}, \quad  \forall n \in [N],
\end{aligned}
\end{eqnarray}
where $d_\text{BS}$ and $d_\text{UE}$ are the inter-element spacing in the BS and UE ULA, respectively. We assume $d_\text{BS}= d_\text{UE} =\frac{\lambda}{2}$, where $\lambda$ is the carrier wavelength. 

%\subsection{Spatial Covariance Estimation}
%The central goal of this work is to estimate the channel covariance matrix $\mathbf{{R}}_h =\mathop{\mathbb{E}}(\mathbf{h}_{t} \mathbf{h}_{t}^\text{H})$ using the finite $T$ under-determined set of measurements $\mathbf{\tilde{y}}_{t,\text{agg}}$. As stated before, this goal can be achieved in at least two possible ways: 
%\begin{enumerate}
%    \item \textit{Indirect Method:} Covariance estimation via the channel estimates -  The key idea is to obtain the channel estimates $\mathbf{\hat{h}}_t, \forall t \in [T]$. Upon obtaining the channel estimates, the covariance can be obtained using the $\mathbf{\hat{R}}_h = \mathbb{E}(\mathbf{\hat{h}}_t\mathbf{\hat{h}}_t^\text{H})$.
%    \item \textit{Direct Method:} Explicit covariance estimation. In cases, where the channel estimates are not required then one can explicitly operate on the  measurement covariance $\mathbf{R}_{y}$ to estimate the channel covariance $\mathbf{\hat{R}}_{h}$. This approach is the direct method which estimates the channel covariance directly as opposed to the former indirect method of obtaining the covariance matrix.
%\end{enumerate}

\section{Sparse Representation and Off-Grid Effects}\label{Sec:Sparse_Representation}
Due to the sparse nature of the spatial channels, researchers often approach the problem of both channel estimation and covariance estimation as a sparse recovery problem and solve using the CS schemes. These CS-based methods are based on the concept of \textit{virtual channel models} \cite{sayeed2002deconstructing}, which provide a virtual angular representation of MIMO channels as discussed next.

\subsection{Virtual Channel: Sparse Representation of the Channel}\label{SubSub:Discretization}

%All these CS-based spatial channel and covariance estimators are derived based on the concept of \textit{virtual channel model} \cite{sayeed2002deconstructing}, which provides a virtual angular representation of the MIMO channels. The virtual channel model describes the channel with respect to fixed basis functions corresponding to the angles from the finite discrete dictionary. In other words, the continuous parameter space of spatial angle-of-arrival (AoA) and angle-of-departure (AoD) are discretized into a finite set of pre-defined spatial angles which focuses on the sparse representation of the MIMO channels using the finite predefined set of spatial angles. The estimation accuracy of CS  methods based on this discretization is limited by the number of points employed in the dictionary. Although this discretization procedure yields state-of-art performance, it has several intrinsic disadvantages \cite{tang2013compressed_off_grid}, including  the \textit{off-grid} effect.

In order to apply the CS schemes, researchers typically adopt a discretization (or gridding) procedure which reduces the continuous parameter space, continuum AoA-AoD tuple ($\theta^\text{rx} , \theta^\text{tx}$) in the interval $\left([ 0, \pi ) \times [ 0, \pi )\right)$, into a set of finite grid points. To denote it mathematically, we consider $\Theta^\text{rx}$ and $\Theta^\text{tx}$ as the set containing the $G_\text{UE}$ and $G_\text{BS}$ finite discretized  grid points in the AoA and AoD domains, respectively. These discretized grid points are chosen such that they satisfy certain properties based on the scheme adapted. The two popular schemes include uniform sampling of the physical and virtual domains, respectively.
\begin{subsubsection}{Uniform Sampling of the Physical Domain}
The classical approach adapted in the literature is to quantize the AoA-AoD domain uniformly \cite{Uniform_Sampling_AoA_AoD,2018_POMP_ICC,tang2019off}. That is, the AoA/AoD tuple $\left(\theta^\text{rx}, \theta^\text{tx}\right)$ appearing in the array response (\ref{Eq:Array_Response}) is uniformly divided in the angle space $\left( [ 0, \pi ) \times [ 0, \pi ) \right)$ as follows: 
\begin{eqnarray}
\begin{aligned}
\Theta^\text{tx} &= \left\{ \bar{\theta}^\text{tx}_{i} = \frac{(i-1)\pi}{G_\text{BS}} \in [0,\pi),  i = 1, 2, \ldots, G_\text{BS} \right\},\\
\Theta^\text{rx} &= \left\{\bar{\theta}^\text{rx}_{i} = \frac{(i-1) \pi}{G_\text{UE}} \in [0,\pi), i = 1, 2, \ldots, G_\text{UE}\right\}.
\end{aligned}
\end{eqnarray}
This procedure causes the terms ($\cos(\theta^{rx}), \cos(\theta^{tx})$) appearing in the array response definition (\ref{Eq:Array_Response}) non-uniformly quantized in the space of $\left([1,-1) \times [1,-1) \right)$ leading to the loss of orthogonality between the discretized virtual channel columns. The natural extension to preserve orthogonality is to uniformly discretize the $\left(\cos(\theta^\text{rx}), \cos(\theta^\text{tx})\right)$ space instead of $\left(\theta^\text{rx}, \theta^\text{tx}\right)$ as presented next. 
\end{subsubsection}

\begin{subsubsection}{Uniform Sampling of the Virtual Domain}
In this approach, the AoA/AoD are taken from a non-uniformly quantized grid such that the $\left(\cos(\theta^\text{rx}), \cos(\theta^\text{tx})\right)$ space appearing in the array response is uniformly quantized \cite{Uniform_Sampling_Cos_Domain,2017_Robust_Beam_Alignment}. The authors in \cite{Uniform_Sampling_Cos_Domain} show that such grids reduce the coherence of the redundant dictionary due to preserving orthogonality which does not hold in the former approach. The grid angles in this approach follow the inverse cosine function as follows
\begin{eqnarray}
\begin{aligned}
\Theta^\text{tx} &= \left\{\bar{\theta}^\text{tx}_{i}: \cos(\bar{\theta}^\text{tx}_{i}) = 1 - \frac{2(i-1)}{G_\text{BS}} \in [1,-1), i = 1, 2, \ldots, G_\text{BS}\right\},\\
\Theta^\text{rx} &= \left\{\bar{\theta}^\text{rx}_{i}: \cos(\bar{\theta}^\text{rx}_{i}) = 1 - \frac{2(i-1)}{G_\text{UE}} \in [1,-1), i = 1, 2, \ldots, G_\text{UE}\right\}.
\end{aligned}
\end{eqnarray}
\end{subsubsection}

%\begin{figure}[!t]
%    \centerline{\includegraphics[scale=0.5]{Uniform_Theta_Physical.eps}
%    \includegraphics[scale=0.5]{Uniform_Theta_Virtual.eps}}
%    \caption{Uniform sampling of $\theta$ domain: Visualization of the grid points in the physical and virtual domains for $G_\text{UE}= G_\text{BS} = 8$.}
%    \label{fig:Theta_Domain}
%\end{figure}

%\begin{figure}[!t]
%    \centerline{\includegraphics[scale=0.5]{Uniform_Cos_Theta_Physical.eps}
%    \includegraphics[scale=0.5]{Uniform_Cos_Theta_Virtual.eps}}
%   \caption{Uniform sampling of $\cos(\theta)$ domain: Visualization of the grid points in the physical and virtual domains for $G_\text{UE}= G_\text{BS} = 8$.}    \label{fig:Cos_Theta_Domain}
%\end{figure}
%Fig. \ref{fig:Theta_Domain} and \ref{fig:Cos_Theta_Domain} provides the visualizations of the grid spacing in the physical and virtual domains for both uniform sampling of $\theta$ and $\cos(\theta)$ domain, respectively.

Here and throughout, we refer to the former scheme as \textit{uniform sampling of $\theta$ domain} and the latter scheme as \textit{uniform sampling of $\cos(\theta)$ domain}. Intuitively, the latter scheme is suitable for models which has significant MPCs in the antenna broadside direction. This is because it has more grid points clustered towards the center which assist to capture the LOS path and MPCs in the broadside direction for the exact sparse representation (higher the grids better the approximation). On the other hand, it might fail to do so for the MPCs that fall away from the center of the broadside as the grid spacing increases away from the center. Contrary to this, the former approach has equal spacing in the physical domain, but as apparent, it creates non-uniform spacing in the virtual space which leads to the loss of orthogonality, and in turn increases in the mutual coherence with the number of grid points. These subtle insights are pivotal in the development of our algorithms discussed in Section \ref{Sec: Proposed Soltuion}. Thus, in our work we adopt both the schemes and validate them using the numerical results. 

\subsection{Sparse Recovery Problem}
Collecting all the array responses corresponding to the angles from the set ($\Theta^\text{rx} , \Theta^\text{tx} $), the array response matrices $\mathbf{A}_\text{UE}(\Theta^\text{rx}) = [\mathbf{a}_\text{UE}(\bar{\theta}_{1}^\text{rx}), \ldots , \mathbf{a}_\text{UE}(\bar{\theta}_{G_\text{UE}}^\text{rx}) ]$ and $\mathbf{A}_\text{BS}(\Theta^\text{tx}) = [\mathbf{a}_\text{BS}(\bar{\theta}_{1}^\text{tx}), \ldots, \mathbf{a}_\text{BS}(\bar{\theta}_{G_\text{BS}}^\text{tx}) ]$ are formed. Using these array response matrices, the channel matrix $\bf{H_t}$ can be represented by the virtual sparse channel (\ref{Virtual_Channel}) which provides a discretized approximation of the channel response
\begin{equation}\label{Virtual_Channel}
    \mathbf{H}_t = \mathbf{A}_\text{UE} \mathbf{H}_{V_t} \mathbf{A}_\text{BS}^\text{H},  \quad \forall t \in [T],
\end{equation}
where $\mathbf{H}_{V_t} \in \mathcal{C}^{G_\text{UE} \times G_\text{BS}}$ is the sparse virtual matrix containing the quantized spatial frequencies. Aided by the sparse virtual representation and vector identity property\footnote{ vec($\bf{ABC}$) = $(\mathbf{C}^T \otimes \mathbf{A}) \text{vec}(\mathbf{B})$.}, the MIMO channel estimation (\ref{y_Equ_Agg1}) is rewritten as a sparse recovery problem
\begin{eqnarray}\label{MMV_Sparse_Probelm}
\begin{aligned}
\mathbf{\tilde{y}}_{t,\text{agg}} &= \mathbf{ \Phi}_{t,\text{agg}} \mathbf{\Psi} \mathbf{h}_{V_t} + \mathbf{\tilde{n}}_{t,\text{agg}},  \quad \forall t \in [T],
\end{aligned}
\end{eqnarray}
where $\mathbf{\Phi}_{t,\text{agg}}\in \mathcal{C}^{M_\text{RF}N_\text{RF} \times M N}$ and $\mathbf{\tilde{n}}_{t,\text{agg}} \in \mathcal{C}^{M_\text{RF}N_\text{RF} \times 1}$ are the combined operation of precoder-combiner and the aggregated noise vector as defined in (\ref{y_Equ_Agg}). The matrix $\mathbf{\Psi} = (\mathbf{A}_\text{BS}^\text{H} \otimes \mathbf{A}_\text{UE}) \in \mathcal{C}^{MN \times G_\text{BS}G_\text{UE}  }$ is the dictionary matrix, where each column of $\mathbf{\Psi}$ denoted as $\Psi_i$ contains the vectorized version of the array response for a specific AoA-AoD grid combination depending on the scheme employed. $\mathbf{h}_{V_t} \in \mathcal{C}^{G_\text{UE}G_\text{BS} \times 1}$ is the vectorized form of $\mathbf{H}_{V_t}$. The setting in (\ref{MMV_Sparse_Probelm}) is the classical joint sparse MMV setup, however, with the time varying sensing matrix. The $\mathbf{h}_{V_t}, \forall t\in [T]$ are assumed to be jointly sparse vectors with the same common support $\mathcal{S} = \text{supp}(\mathbf{h}_{V_t})$ with time-varying weights.

The conventional CS techniques assume that the virtual channel $\mathbf{h}_{V_t}$ is exactly sparse, which is true only when the AoA-AoD tuples are aligned with discretized spatial angles which is an ideal on-grid case. However, the physical AoA-AoD can take any continuous values as defined in model (\ref{H_Equation}), which may not be aligned with any discretized spatial angles causing  off-the-grid effects. These effects violate the sparsity assumption, resulting in performance degradation of CS-based techniques \cite{2018_POMP_ICC, chi2011sensitivity}. Next we discuss this off-grid phenomena and provide an off-grid aware representation in conjunction with the discretized dictionary.  
% Hereafter, $\mathbf{Y} = [\mathbf{\tilde{y}}_{1,\text{agg}}, \ldots, \mathbf{\tilde{y}}_{t,\text{agg}}]  \in \mathcal{C}^{M_\text{RF}N_\text{RF}\times T}$ denotes $\mathbf{\tilde{y}}_{t,\text{agg}}$ stacked in columns.
\subsection{Off-Grid Effect}
The source of the off-grid complication is that the true continuum AoA/AoD tuple $({\theta}^\text{rx},{\theta}^\text{tx})$ may not coincide with one of the predefined discretized grid points in $(\Theta^\text{rx},\Theta^\text{tx})$ as defined in (\ref{Eq:Array_Response}), but may be perturbed away from these grid points with unknown perturbation. This implies that the $\mathbf{h}_{V_t}$ may not be exactly sparse in the assumed basis $\mathbf{\Psi}$, but in the unknown basis $\mathbf{\hat{\Psi}}$. Since it is assumed that the total number of MPCs is $K_L$, there exist $K_L$ columns of $\mathbf{\Psi}$ that needs to be updated adaptively. We denote the indices corresponding to these $K_L$ columns as $\mathcal{K_L}$. At first, we investigate the perturbation mechanism for a single MPC. After we see how to address a single MPC, extending it to accommodate multiple MPCs is relatively straightforward.

%
%= \mathbf{\Psi} \odot \mathbf{P}

Mathematically, the true AoA $\theta_l^\text{rx}$ of the $l^\text{th}$ MPC and the perturbation from the nearest grid point can be related as: $\theta_l^\text{rx} = \bar{\theta}_l^\text{rx} + \delta_l^\text{rx}$, where $\bar{\theta}_l^\text{rx}$ is the grid point that is closest to the true AoA from the set $\Theta^\text{rx}$, and $\delta_l^\text{rx}$ is the perturbation parameter in a bounded AoA space. This bounded space is dependent on the sampling scheme and the number of grid points employed during the creation of a dictionary matrix which is detailed in Section \ref{Sec: Proposed Soltuion}.  A similar relation holds for the true AoD and the AoD perturbation as $\theta_l^\text{rx} = \bar{\theta}_l^\text{rx} + \delta_l^\text{rx}$. The unknown basis for the $l^\text{th}$ MPC can then be related to the nearest discretized grid and perturbation as follows
\begin{align}
{\hat{\Psi}}_l &= \text{vec}\left(\mathbf{a}_\text{UE}(\theta_l^\text{rx})\mathbf{a}_\text{BS}(\theta_l^\text{tx})^\text{H}\right) = \text{vec}\left(\mathbf{a}_\text{UE}(\bar{\theta}_l^\text{rx} + \delta_l^\text{rx})\mathbf{a}_\text{BS}(\bar{\theta}_l^\text{tx} + \delta_l^\text{tx})^\text{H}\right). %\\
%&= \text{vec}(\mathbf{a}_\text{UE}(\bar{\theta}_l^\text{rx}) \mathbf{a}_\text{UE}(\delta_l^\text{rx})\mathbf{a}_\text{BS}(\bar{\theta}_l^\text{tx})^\text{H}\mathbf{a}_\text{BS}(\delta_l^\text{tx})^\text{H}) \\
%& = \text{vec}(\mathbf{a}_\text{UE}(\bar{\theta}_l^\text{rx}) \mathbf{a}_\text{BS}(\bar{\theta}_l^\text{tx})^\text{H})
%\odot 
%\text{vec}(\mathbf{a}_\text{UE}(\delta_l^\text{rx})\mathbf{a}_\text{BS}(\delta_l^\text{tx})^\text{H})\\
%& = {\Psi}_l \odot {P}_l,
\end{align}
%where ${\Psi}_l$ and  $\mathbf{P}_l$ are the vectorized array response corresponding to the discretized grid point closest to the $l^\text{th}$ MPC and the perturbation parameters. The operation $\odot$ denotes the Hadamard product.  Note that all the components of $\mathbf{P}_{\mathcal{K_L}^\mathcal{C}} = 1$.
The unknown basis for all the $K_L$ MPCs can be related as $\mathbf{\hat{\Psi}}_\mathcal{K_L} = [ {\Psi}_{1}, \ldots, {{\Psi}}_{K_L} ] $. If the perturbation parameters can be found then the degradation due to off-grid can be reduced significantly. From this perspective, it becomes clear why capturing the perturbations might be necessary for the optimal sparse representation of the virtual channel model. Thus, the key idea is to solve for the perturbations from the discretized grid points.

\section{Parameter Perturbed Channel Estimation}\label{Sec: Proposed Soltuion}
%\subsection{CS based channel estimation}
In this section, we propose a novel iterative parameter perturbed SOMP (PPSOMP) algorithm for the MMV channel estimation. The proposed algorithm evades the issue arising from the basis mismatch problem by operating on the continuum AoA-AoD space using the mechanism of the controlled perturbation in conjunction with a modified simultaneous OMP (SOMP) framework. The SOMP framework helps to preserves the low computational complexity which is inherent for a greedy solver. Finally, we calculate the spatial covariance matrix  using the channel estimated using the PPSOMP solver. 
%The SOMP performs joint support recovery and is a natural choice because of its low complexity.

We approach the joint problem (\ref{MMV_Sparse_Probelm}) in an iterative fashion within a SOMP framework. The key idea of the PPSOMP solver is as follows: First, we find one of the pre-defined grid points which has highest correlation with the residual measurements and add it to the support set $\mathcal{S}_k$. In so doing, the second crucial step is to invoke the controlled perturbation mechanism to find the perturbations in the AoA/AoD domain for all the points in the support set jointly . 

At any iteration $k$, the measurements $\mathbf{y}_{t,\text{agg}}$ can be decomposed as \cite{2018_POMP_ICC,teke2013perturbed}
\begin{equation}\label{Y_Decomposition}
\mathbf{\tilde{y}}_{t,\text{agg}} = \mathbf{\tilde{y}}_{t,\text{agg}_\perp} + \mathbf{\tilde{y}}_{t,\text{agg}_\parallel}, \quad \forall t \in [T],
\end{equation}
where $\mathbf{\tilde{y}}_{t,\text{agg}_\perp}$ and $ \mathbf{\tilde{y}}_{t,\text{agg}_\parallel}$ are the orthogonal residual and the projection of $\mathbf{y}_{t,\text{agg}}$ onto the span of vectors in the support set $\mathcal{S}_k$ chosen in an iterative fashion. Since the vectors in $\mathcal{S}_k$ are linearly independent, the orthogonal residual in terms of the measurement $\mathbf{y}_{t,\text{agg}}$ and the projection of $\mathbf{\tilde{y}}_{t,\text{agg}_{\parallel}}$, for each snapshot, can be uniquely expressed as
\begin{equation}\label{Eq:Y_Parallel}
\mathbf{\tilde{y}}_{t,\text{agg}_\perp} = \mathbf{{\tilde{y}}}_{t,\text{agg}}   - \mathbf{\Phi}_{t,\text{agg}} \sum_{l=1}^{k} \alpha_{l,t} \mathbf{a}(\theta^\text{rx}_{l},\theta^\text{tx}_{l}),
\end{equation}
where $\mathbf{a}(\theta^\text{rx}_{l} ,\theta^\text{tx}_{l}) = \text{vec}(\mathbf{a}_\text{UE}(\theta^\text{rx}_{l} ) \mathbf{a}_\text{BS}(\theta^\text{tx}_{l})^\text{H})$ denotes the vectorized version of the array response for the AoA-AoD tuple.
% In terms of the perturbations (\ref{Eq:Y_Parallel}) can be expressed as
%\begin{equation}\label{Eq:Y_Parallel_Perturb}
%\mathbf{{\tilde{y}}}_{t,\text{agg}_{\parallel}} = \mathbf{\Phi}_{t,\text{agg}} \sum_{i=1}^{K_L} \alpha_{i,t} %\mathbf{a}(\bar{\theta}^{rx}_i + \delta^{rx}_i ,\bar{\theta}_i^{tx} + \delta^{tx}_i )
%\end{equation}
The goal at each iteration is to choose an initial grid point which minimizes the orthogonal residual as much as possible and this is achieved by the classical projection operation of the SOMP algorithm. 

\subsection{Finding Initial Grid Points}
In the standard SOMP algorithm \cite{SOMP}, the projection step selects a column vector of the sensing matrix that has the largest correlation with the current residual. However, this cannot be directly applied to our system model due to the time-varying sensing matrix. Thus, to adapt to the time-varying system model we modify the projection step as $j^{*} = \arg \underset{j}{\max} \sum_{t=1}^T | {(\mathbf{\Phi}_{t,\text{agg}}\mathbf{\Psi})}_j^T \mathbf{\tilde{y}}_{t,\text{agg}_\perp}|$ as shown in Algorithm \ref{Algorithm:PDSOMP}. A similar escape path is adopted in \cite{park2018spatial}. 

The first implication is that the index $j^\star$ chosen by the projection step indicates the discretized point most correlated to the true AoA-AoD tuple among all the possible discretized AoA/AoD tuple. Intuitively, this step provides the initial grid points ($\bar{\theta}^\text{rx}_l,\bar{\theta}^\text{tx}_l$) from the predefined discretized set ($\Theta^\text{rx},\Theta^\text{tx}$). The second implication is that this allows one to bound the search space for the perturbations ($\delta^\text{rx}_l,\delta^\text{tx}_l$). Rather than searching the entire space, the search space for ($\delta^\text{rx}_l,\delta^\text{tx}_l$) can be reduced to the grid area of the selected grid point. %Specifically, the search space can be bounded as $ \Delta^\text{tx}_\text{LB} \leq \delta^\text{tx}_i \leq \Delta^\text{tx}_\text{UB}$ and $ \Delta^\text{rx}_\text{LB} \leq \delta^\text{rx}_i \leq \Delta^\text{rx}_\text{UB}$. Where $\Delta^\text{tx}_\text{LB}$ and $\Delta^\text{tx}_\text{UB}$ are the lower and upper bound for the perturbation in the AoD space, respectively. Similarly, $\Delta^\text{rx}_\text{LB}$ and $\Delta^\text{rx}_\text{UB}$ for the AoA space. This bounded space depends on the sampling scheme employed during the creation of the dictionary matrix. 

%($\Delta^\text{tx}_\text{UB}=\Delta^\text{tx}_\text{LB}=\Delta $)
For the uniform sampling of $\theta$ scheme, the discretized space is uniform thus the search space for the perturbations can be bounded within $|\delta^\text{rx}_l| \leq \frac{\Delta^\text{rx}}{2}$. Where $\Delta^\text{rx} = \pi/G_\text{BS}$ is the grid resolution. Similarly,  $|\delta^\text{tx}_l| \leq \frac{\Delta^\text{tx}}{2}$. For the uniform sampling of $\cos(\theta)$ scheme, the bounded space for perturbations is non-uniform and is dependent on the chosen initial grid point. This is because of the non-uniform sampling of the physical domain. The lower and
upper bound for the perturbation in the AoD space can then be related as $\Delta_\text{LB}^\text{tx} = (\bar{\theta}_{l}^\text{tx} - \bar{\theta}_{l-1}^\text{tx})/2 $ and $\Delta_\text{UB}^\text{tx} = (\bar{\theta}_{l+1}^\text{tx} - \bar{\theta}_{l}^\text{tx})/2$, where $\bar{\theta}^\text{tx}_{l-1}$ and $\bar{\theta}^\text{tx}_{l+1}$ are the adjacent grid points for the chosen initial grid point, respectively. Similarly, $\Delta^\text{rx}_\text{LB}$ and $\Delta^\text{rx}_\text{UB}$ for the AoA space. The steps of the proposed PPSOMP are detailed in Algorithm \ref{Algorithm:PDSOMP}. 

\begin{algorithm}[!t]
\KwIn{$\mathbf{\tilde{y}}_{t,\text{agg}}\forall t \in [T]$, $\mathbf{\Phi}_{t,\text{agg}}\forall t \in [T]$, $\mathbf{\Psi}$,  $\epsilon$ $\quad$\\ \textbf{Initialization:} $\mathbf{y}_{\perp,t,0}$ = $\mathbf{\tilde{y}}_{t,\text{agg}}$, $\forall t \in [T]$, $\mathbf{\mathcal{S}}_{0} = \{ \}$, $ e  = \sum_{t=1}^T|| \mathbf{y}_{\perp,t}||_2^2$, $k$ = 1.}
\While{$e < \epsilon$}{
%\For{t=1, 2, \ldots, MaxItr}{
{
$j^{*} = \arg \underset{j}{\max} \sum_{t=1}^T | (\mathbf{\Phi}_{t,\text{agg}}\mathbf{\Psi})_j^\text{T} \mathbf{\tilde{y}}_{t,\text{agg}_\perp}|$\\
$\mathcal{S}_{k} = \mathcal{S}_{{k}-1} \cup (\mathbf{\Phi}_{t,\text{agg}}\mathbf{\Psi})_{j^\star}$\\
$({\bm{\alpha}, \bm{\delta}^\text{tx}, \bm{\delta}^\text{rx}} ) = \mathbb{S}(\mathbf{y}_{t,\text{agg}},\mathcal{S}_{k})$\\
$\mathbf{\tilde{y}}_{\perp,t} = \mathbf{\tilde{y}}_{t,\text{agg}} - \mathbf{\Phi}_{t,\text{agg}} \sum_{l=1}^{k}\alpha_{l,t} \mathbf{a}(\bar{\theta}^\text{rx}_l + \delta^\text{rx}_l, \bar{\theta}^\text{tx}_{l} + \delta^\text{tx}_l)$, $\forall t \in [T]$\\
$e = \sum_{t=1}^T || \mathbf{y}_{\perp,t}||_2^2$\\
$k$ = $k$ + 1
}
}
%}
\KwOut{$\mathbf{h}_{t} = \sum_{l=1}^{k} \alpha_{i,t} \mathbf{a}(\bar{\theta}^\text{rx}_l + \delta^\text{rx}_l, \bar{\theta}^\text{tx}_{l} + \delta^\text{tx}_l)$, $\forall t \in [T]$}
\caption{Channel Estimation: PPSOMP - Main Solver} \label{Algorithm:PDSOMP}
\end{algorithm}

\subsection{Finding Perturbations}
For a noiseless condition and under no basis mismatch, the $\mathbf{y}_{t,\text{agg}_\perp}$ would  go to zero after $K_L$ iterations for recovering a $K_L$-sparse vector. However, for off-grid targets and noisy environment, the goal is to reduce the residual term as small as possible and this can be achieved by solving the following joint optimization problem
\begin{eqnarray}\label{MMV_DCOMP}   
\begin{aligned} 
      & \underset{ \{ \alpha_{l,t},\delta^\text{rx}_{l},\delta^\text{tx}_{i} \} }{\text{min}} \sum_{t=1}^T|| \mathbf{\tilde{y}}_{t,\text{agg}} -   \mathbf{\Phi}_{t,\text{agg}} \sum_{l=1}^{k} \alpha_{l,t} \mathbf{a}( \bar{\theta}^\text{rx}_l + \delta^\text{rx}_l, \bar{\theta}^\text{tx}_l + \delta^\text{tx}_l) ||_2^2,\\
       &\quad  \textit{s.t.} \quad  \Delta^\text{tx}_\text{LB} \leq \delta^\text{tx}_l \leq  \Delta^\text{tx}_\text{UB}, \quad  \Delta^\text{rx}_\text{LB} \leq \delta^\text{rx}_l \leq  \Delta^\text{rx}_\text{UB}, \quad \forall l \in [k]. 
\end{aligned}
\end{eqnarray}
The optimization problem (\ref{MMV_DCOMP}) returns the solutions for perturbation parameters $\{ \delta^\text{rx}_l,  \delta^\text{tx}_l\}$, $\forall l \in [k]$ and the weights $\alpha_{l,t}, \forall l \in [l],\forall  t \in [T]$ which is denoted as $(\bm{\alpha, \delta^\text{rx}, \delta^\text{tx}})$.
This procedure is detailed in Algorithm \ref{Algorithm:PDSOMP_Solver}. At the $k^\text{th}$ iteration, starting from the initial grid points provided by the Algorithm \ref{Algorithm:PDSOMP}, the AoA-AoD parameters for all the $k$ MPCs will be  jointly updated within their respective grid regions towards the direction that reduces the sum of residual norms the most.

\begin{algorithm}[!t]
\KwIn{$\mathbf{\tilde{y}}_{t,\text{agg}}\forall t \in [T]$, $\mathbf{\Phi}_{t,\text{agg}}\forall t \in [T]$
\\ \textbf{Initialization:} $p$ = 1, Initial Grid points: $\theta^\text{rx}_{l,p} = \bar{\theta}^\text{rx}_{l}; \forall l \in [k]$, $\theta^\text{tx}_{l,p} = \bar{\theta}^\text{tx}_{l}; \forall l\in [k]$ }
\While{(Until the stopping criterion is met)}{
%\For{t=1, 2, \ldots, MaxItr}{
{
%$\mathbf{\Psi}_{1:k} = [\mathbf{a}(\theta^\text{rx}_{1,p},\theta^\text{tx}_{1,p}), \ldots, \mathbf{a}(\theta^\text{rx}_{k,p},\theta^\text{tx}_{k,p})]$\\
${\Psi}_{l} = \mathbf{a}(\theta^\text{rx}_{l,p},\theta^\text{tx}_{l,p}),\quad \forall l \in [k]$\\
${\alpha}_{l,t} = (\mathbf{\Phi}_{t,\text{agg}}{\Psi}_l)^{\dagger}$ $\mathbf{\tilde{y}}_{t,\text{agg}}, \quad \forall l \in [k], \quad \forall t \in [T]$ \\
$\mathbf{r}_{t,p} = \mathbf{\tilde{y}}_{t,\text{agg}} - \mathbf{\Phi}_{t,\text{agg}} \mathbf{\Psi}_{1,k} \bm{\alpha}_{t}$
\\ $\mathbf{r}_p = \sum_{t=1}^T \mathbf{r}_{t,p}$\\
\text{Update $\mathbf{B}_{p}^\text{rx}$ and $\mathbf{B}_{p}^\text{tx}$ as in (\ref{B_Update_DSOMP})}\\
${\theta^\text{rx}_{l,p+1} = \max{ \{\bar{\theta}^\text{rx}_l -  \Delta_\text{LB}^\text{rx},\min \{\bar{\theta}^\text{rx}_l +\Delta_\text{UB}^\text{rx}, \theta^\text{rx}_{l,p} +  \mu_{p} \mathbb{R}\{\mathbf{B}_{p}^\text{rx} \mathbf{r}_{p} \}\} }\} },\quad \forall l \in [k]$\\
${\theta^\text{tx}_{l,p+1} = \max{ \{\bar{\theta}^\text{tx}_l-\Delta_\text{LB}^\text{tx},\min \{\bar{\theta}^\text{tx}_l + \Delta_\text{UB}^\text{tx}, \theta^\text{tx}_{l,p} +  \mu_{p} \mathbb{R}\{ \mathbf{B}_{p}^\text{tx} \mathbf{r}_{p} \}\} }\} },\quad \forall l \in [k]$\\
$\delta^\text{rx}_l = \theta^\text{rx}_{l,p+1}- \theta^\text{rx}_{l,p} \quad \forall l \in [k]; \quad \delta^\text{tx}_l = \theta^\text{tx}_{l,p+1}- \theta^\text{tx}_{l,p}\quad \forall l \in [k]$\\
$p$ = $p$ + 1
}
}
%}
\KwOut{$\bm{\alpha} = [\alpha_{1,t}, \ldots \alpha_{k,t}], \bm{\delta}^\text{rx}= [\delta^\text{rx}_1, \ldots ,\delta^\text{rx}_k],\bm{\delta}^\text{tx} = [\delta^\text{tx}_1, \ldots ,\delta^\text{tx}_k]$}
\caption{Perturbation Solver $\mathbb{S}$} \label{Algorithm:PDSOMP_Solver}
\end{algorithm}

The AoA/AoD parameters are perturbed as $\theta^\text{tx}_{l,p} =  \bar{\theta}^\text{tx}_{l} + \delta^\text{tx}_{l,p}, \forall l \in [k]$ and $\theta^\text{tx}_{l,p} =  \bar{\theta}^\text{tx}_{l} + \delta^\text{tx}_{l,p}, \forall l \in [k]$, respectively, where $p$ is the perturbation index. At each perturbed point, the weights $\bm{\alpha}_t$ and the perturbations will be updated sequentially in an alternating fashion as shown below
\begin{eqnarray}\label{Alternating_Update_DSOMP}
\begin{aligned}
&\bm{\alpha}_{t,p} = \left[ {\mathbf{a}(\theta^\text{rx}_{1,p},\theta^\text{tx}_{1,p}), \ldots , \mathbf{a}(\theta^\text{rx}_{k,p},\theta^\text{tx}_{k,p}) } \right]^{\dagger} {\tilde{\mathbf{y}}_{t,\text{agg}}},\\
&{\theta^\text{rx}_{l,p+1} = \max{ \{\bar{\theta}^\text{rx}_l -  \Delta_\text{LB}^\text{rx},\min \{\bar{\theta}^\text{rx}_l +\Delta_\text{UB}^\text{rx}, \theta^\text{rx}_{l,p} +  \mu_{p} \mathbb{R}\{ \mathbf{B}_{(l,:)}^\text{rx} \mathbf{r}_{p} \}\} }\} }, \quad \forall l \in [k],\\
&{\theta^\text{tx}_{l,p+1} = \max{ \{\bar{\theta}^\text{tx}_l-\Delta_\text{LB}^\text{tx},\min \{\bar{\theta}^\text{tx}_l + \Delta_\text{UB}^\text{tx}, \theta^\text{tx}_{l,p} +  \mu_{p} \mathbb{R}\{ \mathbf{B}_{(l,:)}^\text{tx} \mathbf{r}_{p} \}\} }\} }, \quad \forall l \in [k],
\end{aligned}
\end{eqnarray}
where $\mu_{p}$ is the step size at the $p^\text{th}$ iteration, $\mathbf{r}_{p} = \sum_{t=1}^T \mathbf{r}_{t,p}$ is the residual update during the $p^\text{th}$ iteration. Note that the bounding of $\theta^\text{rx}_{l,p+1}$ and $\theta^\text{tx}_{l,p+1}$ by the max and min terms at each iteration is essentially the same as bounding the perturbation parameters within the perturbation space. The matrices ${\mathbf{B}^\text{rx}} \in \mathcal{C}^{k \times M_\text{RF}N_\text{RF} }$ and ${\mathbf{B}^\text{tx}}\in \mathcal{C}^{k \times M_\text{RF}N_\text{RF} }$ holds the weighted partial derivatives with respect to the AoA and AoD, respectively, at the $p^\text{th}$ iteration of the parameter point and is mathematically represented as
\begin{eqnarray}\label{B_Update_DSOMP}
\begin{aligned}
&{ \mathbf{B}^\text{rx} =  \left[{ \left(\sum_{t=1}^T\alpha_{1,t} \Phi_{t,\text{agg}}\frac{\partial \mathbf{a}(\theta^\text{rx}_{1},\theta^\text{tx}_{1})}{\partial \theta^\text{rx}_{1}}\right), \ldots , \left( \sum_{t=1}^T\alpha_{k,t} \Phi_{t,\text{agg}} \frac{\partial \mathbf{a}(\theta^\text{rx}_{k},\theta^\text{tx}_{k})}{\partial \theta^\text{rx}_{k}} \right)} \right]^\text{H}},\\% \quad \bf{B_{t,l}^{rx} = \delta^{rx} \alpha_{t,l} B_{l}^{rx} };\\
&{ \mathbf{B}^\text{tx} =  \left[{ \left(\sum_{t=1}^T\alpha_{1,t} \Phi_{t,\text{agg}} \frac{\partial \mathbf{a}(\theta^\text{rx}_{1},\theta^\text{tx}_{1})}{\partial \theta^\text{tx}_{1}}\right), \ldots , \left(\sum_{t=1}^T\alpha_{k,t}\Phi_{t,\text{agg}} \frac{\partial \mathbf{a}(\theta^\text{tx}_{k},\theta^\text{tx}_{k})}{\partial \theta^\text{tx}_{k}}\right)} \right]^\text{H}}.
\end{aligned}
\end{eqnarray}

At this point, some remarks on Algorithm \ref{Algorithm:PDSOMP} and Algorithm \ref{Algorithm:PDSOMP_Solver} are in order
\begin{remark}
The PPSOMP main solver in Algorithm \ref{Algorithm:PDSOMP} is generalized for the MMV setup and reduces to the PPOMP SMV work \cite{2018_POMP_ICC} when $T$ = 1. The step 2 in Algorithm \ref{Algorithm:PDSOMP} is the greedy projection/selection step which chooses the initial grid points for the perturbation solver in Algorithm \ref{Algorithm:PDSOMP_Solver}. The remaining steps are self explanatory and are repeated until the stopping criterion is met. The critical advantage of Algorithm \ref{Algorithm:PDSOMP} is that it preserves the low-complexity of the greedy approach and provides the initial grid points for each MPC in an iterative fashion.
\end{remark}
\begin{remark}The Perturbation solver $\mathbb{S}$ in Algorithm 2 uses gradient based updates to jointly find the perturbation parameters $(\bm{\delta}^\text{rx},\bm{\delta}^\text{tx})$ and $\bm{\alpha}$ of the MPCs that reduces the residual the most. This is detailed in Algorithm \ref{Algorithm:PDSOMP_Solver}. The AoA/AoD parameters are jointly updated within their respective grid region and is made sure not to cross the upper and lower bound of the grid points (steps 7 and 8). Further, it is important to note that the perturbations solved by Algorithm \ref{Algorithm:PDSOMP_Solver} is valid only if Algorithm \ref{Algorithm:PDSOMP} finds the correct support set. 
\end{remark}

\begin{remark} The convergence of the perturbation solver $\mathbb{S}$ depends on the choice of $\mu_p$ and can further be improved by using acceleration schemes based on conjugate gradient methods \cite{Conjugate_Gradient}, Newton and Quasi-Newton methods \cite{Newton,Quasi_Newton} and so on. However, in this work, we restrict our discussion to the gradient descent scheme.
\end{remark}

%In some sense, what we describe in Algorithm 1 and 2 is a way to traverse to the true AoA-AoD by finding the perturbation in the continuum space provided the initial grid points. Since PPSOMP operates in the continuum space it combats the off-grid issues arising from the discretization procedure. We validate our theory and claims in the numerical section which confirms the superiority over just the discretization techniques alone.

\subsection{Covariance Estimation via the CS based Channel Estimation}
The channel estimates $\mathbf{h}_{t}, \forall t \in [T]$ obtained from the PPSOMP algorithm allows us to calculate the channel covariance matrix ${\hat{\mathbf{R}}_{h}} =\mathbb{E}(\hat{\mathbf{h}}_t \hat{\mathbf{h}}_t^\text{H})$. As apparent, the quality of this indirect covariance estimate scheme depends on the quality of channel estimates obtained across all the snapshots. As stated before, when the channel estimates are not required then one can explicitly estimate the covariance matrix directly. This scheme is presented up next. 

\section{Parameter Perturbed Covariance Estimation}\label{Sec:Explicit_Covariance_Estimation}
The covariance matrix ${{\mathbf{R}}_{h}}$ can be explicitly estimated as opposed to the indirect approach presented in the previous section. This can be made possible by relating the channel covariance ${{\mathbf{R}}_{h}} = \mathbb{E}(\mathbf{h}_t\mathbf{h}_t^\text{H})$ and the covariance of the measurements ${\mathbf{R}_{{y}}} = \mathbb{E}(\tilde{\mathbf{y}}_{t,\text{agg}}\tilde{\mathbf{y}}_{t,\text{agg}}^\text{H})$. However, due to the time-varying sensing matrices ${\mathbf{\Phi}_{t,\text{agg}}}$ the covariance matrix ${{\mathbf{R}}_{h}}$ cannot be explicitly written as a function of the covariance of measurement ${{\mathbf{R}}_{y}}$ but can only be related via the per snapshot covariance matrix as follows:
\begin{eqnarray}\label{Eq:Covariance_Problem}
\begin{aligned}
{\mathbf{R}_{\tilde{y}_{t,\text{agg}}}} &= {\tilde{\mathbf{y}}_{t,\text{agg}} \tilde{\mathbf{y}}_{t,\text{agg}}^\text{H}}, \quad t \in [T],\\
{\mathbf{R}_{\tilde{y}_{t,\text{agg}}}} &=  {\mathbf{\Phi}_{t,\text{agg}} {\mathbf{R}_{h_t}} \mathbf{\Phi}_{t,\text{agg}}^\text{H}} + {\mathbf{N}_{t,\text{agg}}}+  {\mathbf{Z}_{t,\text{agg}} }, \quad t \in [T],
\end{aligned}
\end{eqnarray}
%{\mathbf{\Psi} \mathbf{R}_{h_{V_t}}  \mathbf{\Psi}^\text{H}}
where ${\mathbf{R}_{\tilde{y}_{t,\text{agg}}}}$ is the per snapshot covariance matrix of the measurements $\tilde{\mathbf{y}}_{t,\text{agg}}$. The per snapshot channel covariance ${\mathbf{R}_{h_{t}}}$ is defined as ${\mathbf{h}_{t} \mathbf{h}_{t}^\text{H}}$. The matrix ${\mathbf{N}_{t,\text{agg}}} = { \tilde{\mathbf{n}}_{t,\text{agg}} \tilde{\mathbf{n}}_{t,\text{agg}}^\text{H} }$ and ${\mathbf{Z}_t} = 2 {\mathbf{\Phi}_{t,\text{agg}} {\mathbf{h}_{t}} \tilde{\mathbf{n}}_{t,\text{agg}}^\text{H}}$ are the per snapshot noise and zero mean signal-noise cross terms, respectively. The zero mean is due to the fact that the AWGN noise and the ${\mathbf{h}_{t}}$ are assumed to be independent with zero mean, respectively. Hereafter, the combined effect of the noise and the signal-noise term is denoted as ${\mathbf{E}_{t,\text{agg}}}$. Note that by construction, all the covariance matrices are inherently Hermitian in nature. Finally, the channel covariance can be obtained as $\mathbf{R}_{h} = \frac{1}{T}\sum_{t=1}^T \mathbf{R}_{h_t}$. 

With the above notations and the aid of virtual channel representation, the formulation in (\ref{Eq:Covariance_Problem}) can be rewritten linearly as
%Further, the signal-noise cross term diminishes with the increase in the number of snapshots resulting in negligible contribution to (\ref{Eq:Covariance_Problem}) and is treated as a residual term in the formulation.
\begin{eqnarray}\label{Reformulated_Covariance_Problem}
\begin{aligned}
{\mathbf{R}_{\tilde{y}_{t,\text{agg}}}} &=  {\mathbf{\Phi}_{t,\text{agg}} \mathbf{\Psi} \mathbf{R}_{h_{v_t}}  \mathbf{\Psi}^\text{H} \mathbf{\Phi}_{t,\text{agg}}^\text{H}} +   {\mathbf{E}_{t,\text{agg}}} , \quad \forall t \in [T],
\end{aligned}
\end{eqnarray}
where $\mathbf{R}_{h_{v_t}}$ are the sparse Hermitian matrices sharing the same support set across all the snapshots. The goal now would be to recover per snapshot virtual covariance matrix ${\mathbf{R}_{h_{v_t}}, \forall t \in [T]}$ using per snapshot covariance matrix of the measurements ${\mathbf{R}_{\hat{y}_{t,\text{agg}}}}, \forall t \in [T]$. Upon obtaining the ${\hat{\mathbf{R}}_{h_{v_t}}}, \forall t \in [T]$ the original channel covariance matrix ${\hat{\mathbf{R}}_{h}}$ can then be obtained by the following relation: ${\hat{\mathbf{R}}_h} = \mathbf{\Psi}  \left(\frac{1}{T}\sum_{t=1}^T \hat{\mathbf{R}}_{h_{v_t}} \right) \mathbf{\Psi}^\text{H}$. Note that the sparse virtual covariance matrix estimation problem in (\ref{Reformulated_Covariance_Problem}) can be reduced to MMV vector type recovery by using the vector identity property similar to the previous approach. However, this vectorized approach would fail to exploit the inherent Hermitian structure of the covariance matrix which can be exploited further to improve the covariance estimation performance. A similar approach is adopted in \cite{park2018spatial} which formulates the sparse covariance estimation as the following optimization problem 
\begin{eqnarray}\label{PPOMP_DCOMP_Opt_Problem}
\begin{aligned}
 \underset{ \{ {\mathbf{R}_{h_{v_t}}} \} }{\text{min}} & \quad \frac{1}{T} \sum_{t=1}^T  ||{\mathbf{R}_{\hat{y}_{t,\text{agg}}}}- {\mathbf{\Phi}_{t,\text{agg}} \mathbf{\Psi} \mathbf{R}_{h_{v_t}} \mathbf{\Psi}_{t,\text{agg}}^\text{H}  \mathbf{\Phi}^\text{H}}||_2^2,\\ 
 \textit{s.t.}& \quad ||\mathbf{R}_{h_{v_t}}||_\text{lattice,0} \leq K_L,
\end{aligned}
\end{eqnarray}
where $||\mathbf{R}_{h_{v_t}}||_\text{lattice,0}$ = $|\cup_{i} \text{supp}([\mathbf{R}_{h_{v_t}}]_{:,i})$ $\cup_{j} \text{supp}([\mathbf{R}_{h_{v_t}}]_{j,:})|$. The above formulation results in disadvantages of twofold: Firstly, in practice, it is difficult to know the total number of MPCs ($K_L$) apriori. Even with the exact knowledge of $K_L$, the number of non-zero components in the ${\mathbf{R}_{h_{v_t}}}$ cannot be $K_L$ because of the basis-mismatch problem. To solve the above mentioned optimization problem, we adopt a similar approach as PPSOMP algorithm with controlled perturbation mechanism which we refer as Parameter Perturbed Covariance OMP (PPCOMP). The PPCSOMP peculiarity lies in considering the covariance space and is designed to exploit the Hermitian property of a covariance, where the diagonal entries (real) are representative of the common MMV support and the off-diagonal are complex conjugates. This structure helps in reducing the number of operations which will become clear shortly.

To adapt the perturbation mechanism to the covariance estimation problem, we rewrite the objective function in (\ref{PPOMP_DCOMP_Opt_Problem}) in terms of the perturbation parameters as
\begin{eqnarray}\label{PPOMP_DCOMP_Opt_Problem_Reform}
\begin{aligned}
\underset{ \{ \Gamma_{l,q,t}, \delta^{rx}_{l}, \delta^{tx}_{l} \} }{\text{min}} & \quad \frac{1}{T} \sum_{t=1}^T  ||{\mathbf{R}_{\tilde{y}_{t,\text{agg}}}}- {   \sum_{l=1}^{K_L} \sum_{q=1}^{K_L} { \Gamma_{l,q,t} \mathbf{\Psi}_{t,\text{agg}} \mathbf{A}_{res}([\theta^\text{rx}_{l}, \theta^\text{tx}_{l}],[\theta^\text{rx}_{q}, \theta^\text{tx}_{q}]) \mathbf{\Psi}_{t,\text{agg}}^\text{H}}}||_F^2,\\ 
\end{aligned}
\end{eqnarray}
where $\mathbf{R}_{h_t} = {\mathbf{h}_t \mathbf{h}_t^\text{H}} =  \sum_{l}^{K_L} \sum_{q}^{K_L} { \Gamma_{l,q,t} \mathbf{A}_{res}([\theta^\text{rx}_{l}, \theta^\text{tx}_{l}],[\theta^\text{rx}_{q}, \theta^\text{tx}_{q}])}$, and ${\Gamma_{l,q,t}} = \alpha_{l,t}\alpha_{q,t}^*$ is the cross term gain between the $l^\text{th}$ and $q^\text{th}$ MPCs at the $t^\text{th}$ snapshot. Further, ${\mathbf{A}_\text{res}([\theta^\text{rx}_{l}, \theta^\text{tx}_{l}],[\theta^\text{rx}_{q}, \theta^\text{tx}_{q}])} = {\mathbf{a}_\text{res}(\theta^\text{rx}_{l}, \theta^\text{tx}_{l})} {\mathbf{a}_\text{res}(\theta^\text{rx}_{q}, \theta^\text{tx}_{q})}^\text{H}$ with $\theta_l^\text{rx} = \bar{\theta}^\text{rx}_l + \delta^\text{rx}_l$ and $\theta_i^\text{rx} = \bar{\theta}^\text{tx}_l + \delta^\text{tx}_l$. Note that (\ref{PPOMP_DCOMP_Opt_Problem_Reform}) is the reformulation of (\ref{PPOMP_DCOMP_Opt_Problem}) in terms of ${\mathbf{R}_{h_t}}$. However, replacing the definition of ${\mathbf{R}_{h_t}}$ is pivotal for the development of PPCOMP solver which follows up next.

\begin{algorithm}[!b]
\KwIn{${\tilde{\mathbf{y}}_{t,\text{agg}}}$, $\forall t \in [T]$, ${\mathbf{\Phi}_{t,\text{agg}}}$, $\forall t \in [T]$, $\epsilon$ 
\\ \textbf{Initialization:} ${\mathbf{R}_{\tilde{y}_{t,\text{agg}_{\perp}}}} = {\mathbf{R}_{\tilde{y}_{t,\text{agg}}}}$ ${\mathcal{S}_{0}} = \{ \}$, $ e  = \sum_{t=1}^T || \mathbf{R}_{\tilde{y}_{t,\text{agg}_{\perp}}}||_F^2$, $k$ = 1.}
\While{$e < \epsilon$}{
%\For{t=1, 2, \ldots, MaxItr}{
{
$j^{\star} = \arg \underset{j}{\max} \sum_{t=1}^T | (\mathbf{\Phi}_{t,\text{agg}}\mathbf{\Psi})_j^\text{H} {\mathbf{R}_{\tilde{y}_{t,\text{agg}_{\perp}}}}(\mathbf{\Phi}_{t,\text{agg}}\mathbf{\Psi})_j|$\\
$\mathcal{S}_{k} = \mathcal{S}_{k-1} \cup j^\star$\\
$({\bm{\Gamma}, \bm{\delta}^\text{rx}, \bm{\delta}^\text{rx}} ) = \mathbb{S}(\mathbf{R}_{\tilde{y}_{t,\text{agg}}},\mathcal{S}_k)$\\
$\mathbf{R}_{\tilde{y}_{t,\text{agg}_{\perp}}} = {\mathbf{R}_{\tilde{y}_{t,\text{agg}}}} - \scriptstyle{{   \sum_{l=1}^{k} \sum_{q=1}^{k}  { \Gamma_{l,q,t} \mathbf{\Phi}_{t,\text{agg}} \mathbf{A}_{res}([\bar{\theta}^\text{rx}_{l}+\delta^\text{rx}_{l}, \bar{\theta}^\text{tx}_{l} +\delta^\text{tx}_l],[\bar{\theta}^\text{rx}_{q}+\delta^\text{rx}_{q}, \hat{\theta}^\text{tx}_{q} + \delta^\text{rx}_{q}]) \mathbf{\Phi}_{t,\text{agg}}^\text{H}}}}$\\
$e$ = $\sum_{t=1}^T || \mathbf{R}_{y_{t,\text{agg}_\perp}}||_F^2$\\
$k = k + 1$
}
}
%}
\KwOut{$\mathbf{R}_{h_t} = {   \sum_{l=1}^{k} \sum_{q=1}^{k}  { \Gamma_{l,q,t} \mathbf{\Phi}_{t,\text{agg}} \mathbf{A}_{res}([\bar{\theta}^\text{rx}_{l}+\delta^\text{rx}_{l}, \bar{\theta}^\text{tx}_{l} +\delta^\text{tx}_l],[\bar{\theta}^\text{rx}_{q}+\delta^\text{rx}_{q}, \hat{\theta}^\text{tx}_{q} + \delta^\text{rx}_{q}]) \mathbf{\Phi}_{t,\text{agg}}^\text{H}}}$}
\caption{Covariance Estimation: PPCOMP - Main Solver} \label{Algorithm:PDCOMP}
\end{algorithm}

Similar to previous section, we solve the optimization problem in (\ref{PPOMP_DCOMP_Opt_Problem}) in a greedy fashion, where we split the problem into finding the initial grid points for each MPC and perturbing the MPCs. The initial grid points are provided in an iterative manner by the projection step in the main solver in Algorithm 3. The notable change in the projection step compared to the PPSOMP is the use of quadratic forms instead of the linear forms to accommodate the measurement covariance \cite{park2018spatial} as shown in Algorithm 3.

At each iteration $k$, provided the initial grid points, the optimization problem in (19) reduces to solving jointly for the $k$ perturbed parameters of the MPCs AoA-AoD and the cross-term gains $\Gamma_{l,q,t}, \forall l\in [k], \forall q\in [k], \forall t \in [T]$ as defined in (\ref{PPOMP_DCOMP_Opt_Problem_Reform}). The procedure to obtain these steps are detailed in Algorithm \ref{Algorithm:PDCOMP_Solver}. At this point, some remark on Algorithm 4 are in order

\begin{remark}
Due to the Hermitian structure, the cross-terms $\Gamma_{l,q,t},\forall l \in [k],\forall q \in [k], \forall t \in [T]$ are only evaluated for $q\geq l$ terms (step 3). The terms $\Gamma_{l,q,t}, q< l = \Gamma_{l,q,t}^\text{H}$, thus saving the computational complexity exploiting the inherent Hermitian property of the covariance matrix.
\end{remark}
\begin{algorithm}[!b]
\KwIn{${\mathbf{\Phi}_{t,\text{agg}}}$, $\forall t \in [T]$, ${\mathbf{R}_{\tilde{y}_{t,\text{agg}}}}$, $\forall t \in [T]$, $p=1$ \\
\hspace{1.2cm} Initial Grid points: $\theta^\text{rx}_{l,p} = \theta^\text{rx}_{l}$, $\forall l \in [k]$, $\theta^\text{tx}_{l,p} = \theta^\text{tx}_{l}$, $\forall l \in [k]$ }
\While{(Until the stopping criterion is met)}{
%\For{t=1, 2, \ldots, MaxItr}{
{
\For{l}{
\For{$q\geq l$}{
$\Gamma_{l,q,t} =  (\mathbf{\Phi}_{t,\text{agg}} \mathbf{a}_\text{res}(\theta^\text{rx}_l,\theta^\text{tx}_l) )^\dagger {\mathbf{R}_{\tilde{y}_{t,\text{agg}}}} \left( (\mathbf{\Phi}_{t,\text{agg}} \mathbf{a}_\text{res}(\theta^\text{rx}_q,\theta^\text{tx}_q) )^\dagger\right)^\text{H}$
}
}
$\Gamma_{l,q,t} = \Gamma_{q,l,t}^*, \quad \forall q < l$\\
$\mathbf{R}_{\tilde{y}_{t,\text{agg}_\perp,p}}= \mathbf{R}_{\tilde{y}_{t,\text{agg}}} - \Psi_{t,\text{agg}} \sum_{l=1}^{k}\sum_{q=1}^{k} \Gamma_{l,q,t} \mathbf{A}_\text{res}([\theta^\text{rx}_l,\theta^\text{tx}_l],[\theta^\text{rx}_q,\theta^\text{tx}_q])$, $\forall t \in [T]$\\
$\mathbf{R}_{p}  = \sum_{t=1}^T\mathbf{R}_{\tilde{y}_{t,\text{agg}_\perp,p}}$\\
\text{Update $\mathbf{B}^\text{rx}$ and $\mathbf{B}^\text{tx}$ as in (\ref{B_Update_DCOMP})}\\
${\theta^\text{rx}_{l,p+1} = \max{\{\bar{\theta}_l^\text{rx} -\Delta_\text{LB}^\text{rx},\min \{\bar{\theta}_l^\text{rx} +\Delta_\text{UB}^\text{rx}, \theta^\text{rx}_{l,p} +  \mu_{p} \mathbb{R}\{ \mathbf{B}_{(l,:)}^\text{rx} \text{vec}(\mathbf{R}_{p}) \}\} }\} }$\\
${\theta^\text{tx}_{l,p+1} = \max{ \{ \bar{\theta}^\text{tx}_l - \Delta_\text{LB}^\text{tx},\min \{  \bar{\theta}^\text{tx}_l +\Delta_\text{UB}^\text{tx}, \theta^\text{tx}_{l,p} +  \mu_{p} \mathbb{R}\{ \mathbf{B}_{(l,:)}^\text{tx} \text{vec}(\mathbf{R}_{p}) \}\} }\} }$\\
$\delta^\text{rx}_l = \theta^\text{rx}_{l,p+1}- \theta^\text{rx}_{l,p} \quad \forall l \in [k]; \quad \delta^\text{tx}_l = \theta^\text{tx}_{l,p+1}- \theta^\text{tx}_{l,p}\quad \forall l \in [k]$\\
$p$ = $p$ + 1
}
}
%}
\KwOut{$\bm{\Gamma} =[\Gamma_{1,1}, \ldots, \Gamma_{k,k}], \bm{\delta}^\text{tx} =[\delta_{1}^\text{tx}, \ldots, \delta_{k}^\text{tx}], \bm{\delta}^\text{rx} =[\delta_{1}^\text{rx}, \ldots, \delta_{k}^\text{rx}]$}
\caption{Covariance Estimation: PPCOMP - Perturbation Solver} \label{Algorithm:PDCOMP_Solver}
\end{algorithm}

\begin{remark}At each iteration $k$, the AoA-AoD parameters are perturbed within their grid regions towards the direction that reduces the norm of the residual measurement covariance the most (step 8 in Algorithm 4). At the $p^\text{th}$ perturbation iteration, the AoA/AoD parameters are perturbed as $\theta^\text{tx}_{l,p} =  \bar{\theta}^\text{tx}_{l} + \delta^\text{tx}_{l,p}$ and $\theta^\text{tx}_{l,p} =  \bar{\theta}^\text{tx}_{l} + \delta^\text{tx}_{l,p}$, where $p$ is the perturbation index.
\end{remark}

At each perturbed point, the weights $\bm{\Gamma}$ and the perturbations will be updated sequentially in an alternating fashion as shown in steps 2 through 5 of Algorithm \ref{Algorithm:PDCOMP_Solver}. The matrices ${\mathbf{B}^\text{rx}} \in \mathcal{C}^{k \times (M_\text{RF}N_\text{RF} )^2}$ and ${\mathbf{B}^\text{tx}}\in \mathcal{C}^{k \times (M_\text{RF}N_\text{RF} )^2}$ holding the weighted partial derivatives with respect to the AoA and AoD, respectively, are mathematically defined as follows:
\begin{eqnarray*}\label{B_Update_DCOMP}
\begin{aligned}
&\scalebox{.9}{$ \mathbf{b}_l^\text{rx} =\left[  \left(\sum_{t=1}^T \sum_{q=1}^k \Gamma_{l,q,t}\right) \text{vec}\left(\Phi_{t,\text{agg}}\frac{\partial \mathbf{A}_\text{res}([\theta^\text{rx}_{l},\theta^\text{tx}_{l}],[\theta^\text{rx}_{q},\theta^\text{tx}_{q}])}{\partial \theta^\text{rx}_{1}}\Phi_{t,\text{agg}}^\text{H}\right)\right]; \quad \mathbf{B}^\text{rx} = \left[ \mathbf{b}_1^\text{rx}, \ldots,\mathbf{b}_k^\text{rx}\right],$}\\%, \ldots ,  \left(\sum_{t=1}^T \sum_{l,q=1}^k \Gamma_{l,q,t}\right) \text{vec}\left(\frac{\partial \mathbf{A}_\text{res}([\theta^\text{rx}_{1},\theta^\text{tx}_{1}],[\theta^\text{rx}_{1},\theta^\text{tx}_{1}])}{\partial \theta^\text{rx}_{l}}\right)\right]^\text{H}$},\\% \quad \bf{B_{t,l}^{rx} = \delta^{rx} \alpha_{t,l} B_{l}^{rx} };\\
&\scalebox{.9}{$ \mathbf{b}_l^\text{tx} =\left[  \left(\sum_{t=1}^T \sum_{q=1}^k \Gamma_{l,q,t}\right) \text{vec}\left(\Phi_{t,\text{agg}}\frac{\partial \mathbf{A}_\text{res}([\theta^\text{rx}_{l},\theta^\text{tx}_{l}],[\theta^\text{rx}_{q},\theta^\text{tx}_{q}])}{\partial \theta^\text{tx}_{1}}\Phi_{t,\text{agg}}^\text{H}\right)\right]; \quad \mathbf{B}^\text{tx} = \left[ \mathbf{b}_1^\text{tx}, \ldots,\mathbf{b}_k^\text{tx}\right].$}
%&\scalebox{.95}{$ \mathbf{B}_{i,l}^\text{tx} =  \left[ \left(\sum_{t=1}^T \sum_{i=1}^{K_L} \sum_{j=1}^{K_L} \Gamma_{i,j,t}\frac{\partial \mathbf{A}_\text{res}([\theta^\text{rx}_{1},\theta^\text{tx}_{1}],[\theta^\text{rx}_{1},\theta^\text{tx}_{1}])}{\partial \theta^\text{tx}_{1}}\right), \ldots , \left(\sum_{t=1}^T \sum_{i=1}^{K_L} \sum_{j=1}^{K_L} \Gamma_{i,j,t}\frac{\partial \mathbf{A}_\text{res}([\theta^\text{rx}_{1},\theta^\text{tx}_{1}],[\theta^\text{rx}_{1},\theta^\text{tx}_{1}])}{\partial \theta^\text{tx}_{i}}\right)\right]^\text{H}.$}
\end{aligned}
\end{eqnarray*}

\section{Numerical Results}\label{sec:results}
In this section, we demonstrate the efficacy of our proposed methods using Monte Carlo simulations. We consider an mmWave MIMO network with $M = 16$ and $N=8$ antennas at the BS and UE, respectively. We assume the channel contains a total of 8 MPCs with the number of clusters $K = 4$ and the number of MPCs per each cluster $L=2$ as found in \cite{ChannelModel_Rappaport}. In particular, the AoA-AoD tuple are not assumed to be on the grids but can take any continuous value in its domain. In specific, the AoA/AoD centers $\theta_{k}^\text{rx}$ and $\theta_{k}^\text{tx}$ are chosen randomly in the interval of $[0,\pi]$. The AoA-AoD angular dispersion is fixed as $\sigma_\text{AS}^\text{AoA} = \sigma_\text{AS}^\text{AoD} = 20^\circ$ \cite{Propogation_Statistics_Chethan_2018}. The complex gain $\alpha_{k,l,t}$ and the noise vector $\mathbf{n}_{t,s}$ are modeled as $i.i.d$ random variables with the complex Gaussian distribution, $\alpha_{k,l,t} \sim \mathcal{CN}(0,1)$ and $\mathbf{n}_{t,s} \sim \mathcal{CN}(0,\sigma_n^2\mathbb{I}_N)$, respectively. Further, the number of grid points for both the AoA/AoD space are chosen to be $G_\text{BS} = G_\text{UE} = 16$ and the stopping criterion parameter $\epsilon$ is chosen to be $10^{-2}$. 

We compare the performance of our proposed algorithms against the benchmark algorithms dynamic SOMP (DSOMP) and covariance OMP (COMP) proposed in \cite{park2018spatial}. All the results presented in this section unless mentioned otherwise are obtained with the above mentioned setting and are averaged over 100 independent trials.

\subsection{Performance Evaluation Metrics}
In the following sections, we evaluate the performance of the proposed algorithms based on two important metrics. The channel estimation algorithms are evaluated based on the \textit{normalized mean square error (NMSE)} metric, defined as = $\mathbb{E}\left(\frac{||\mathbf{H} - \hat{\mathbf{H}} ||_F^2}{||\mathbf{H}||_F^2}\right)$ which we denote as \textit{NMSE-H}. The covariance estimation algorithms are mainly evaluated based on the \textit{relative efficiency metric} as adopted in \cite{Relative_Metric_Eta_Caire,park2017spatial}, which is defined as $\eta = \frac{\mathbf{U}_{\hat{\mathbf{R}}_{h}}^\text{H} \mathbf{R}_h \mathbf{U}_{\hat{\mathbf{R}}_h}}{\mathbf{U}_{\mathbf{R}_h}^\text{H} \mathbf{R}_h \mathbf{U}_{\mathbf{R}_h}} \in [0,1]$. Here $\mathbf{R}_h$ and $\hat{\mathbf{R}}_h$ are the true covariance and the estimated covariance matrix, respectively, while, $\mathbf{U}_{\mathbf{R}_H}$ and $\mathbf{U}_{\hat{\mathbf{R}}_H}$ are the matrices containing the singular vectors corresponding to the singular values of the true covariance and estimated covariance matrices, respectively. Intuitively, $1 -\eta$ denotes the fraction of signal power lost due to the mismatch between the optimal beamformer and its estimate \cite{Relative_Metric_Eta_Caire}. Thus, higher the $\eta$, better are the obtained estimates. We also show the NMSE between the real covariance matrix and estimated covariance as adopted in \cite{mendez2015adaptive}. The \textit{NMSE-Covariance (NMSE-C)} is defined as $\mathbb{E}\left( \frac{||\hat{\mathbf{R}}_h - \mathbf{R}_h||_F^2}{||\mathbf{R}_h||_F^2}\right)$.

\subsection{Channel Estimation: Performance of PPSOMP}
	\begin{figure}[t!]
		\centerline{\includegraphics[scale=0.6]{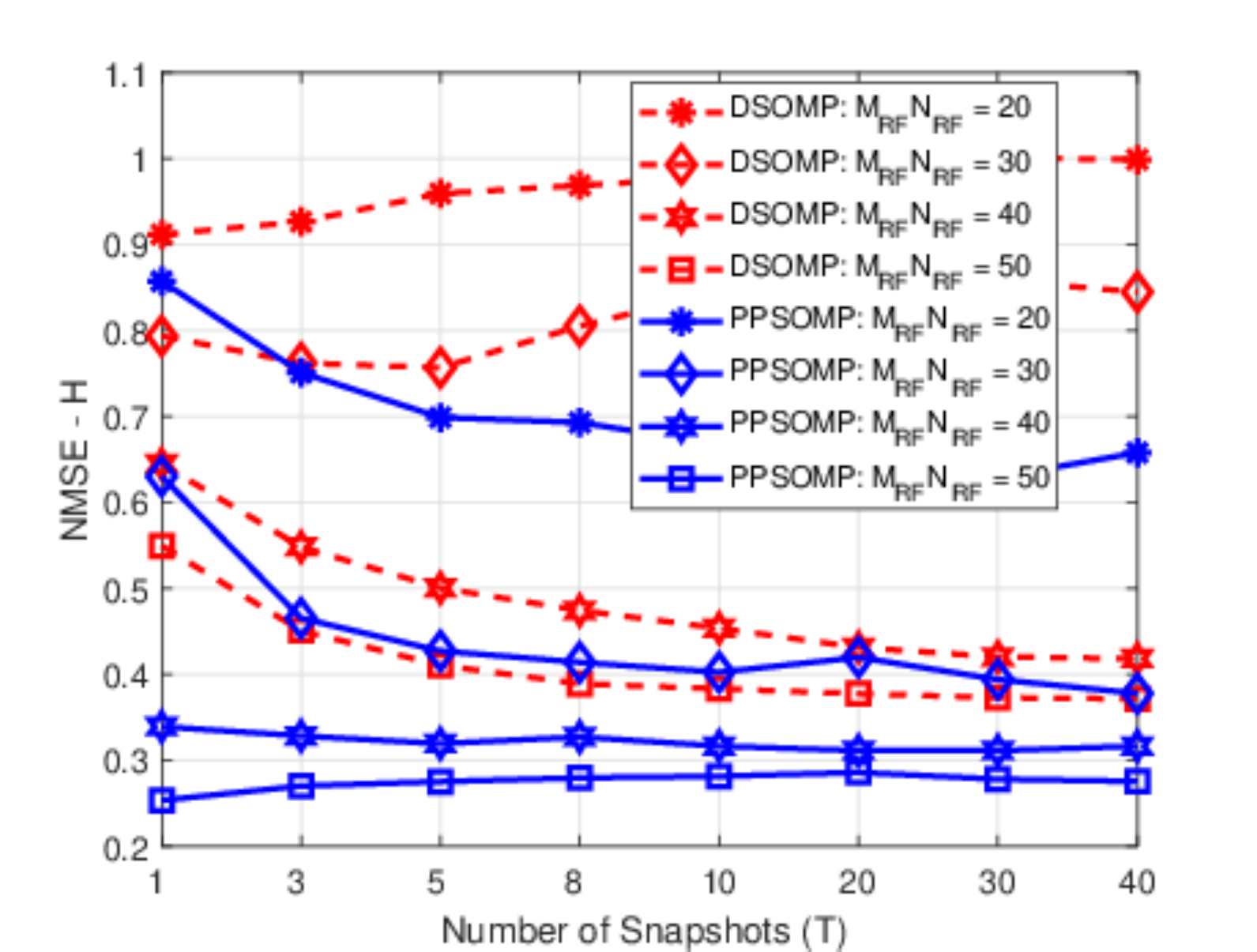}}\vspace{-2mm}
		\caption{Comparison of channel estimation error reconstruction performance versus the number of snapshots ($M_{RF}N_{RF} = \{20, 30 ,40, 50\}$, SNR = 10 dB, and uniform sampling of $\cos(\theta)$ domain).}
		\label{Fig:Channel_Estimation_NMSE_H}
	\end{figure}
Fig. \ref{Fig:Channel_Estimation_NMSE_H} compares the performance of channel estimation algorithms in terms of the NMSE-H for four different levels of measurement numbers $M_{RF}N_{RF}$ = 20, 30, 40, and 50 at an SNR of 10 dB. At low measurement regime ($M_{RF}N_{RF}$ = 20 and 30), the DSOMP performs worse with NMSE-H increasing with the number of snapshots. This performance degradation is exacerbated with number of snapshots as it picks the wrong support and the error gets accumulated with the increase in the number of snapshots. The possible explanation is that the number of measurements on average is lesser than the sparsity level in the virtual channel representation. However, this trend disappears for the DSOMP beyond $M_{RF}N_{RF}$ = 40 implying the number of measurements are adequate. The proposed PPSOMP performance is better than the DSOMP algorithms at all tested cases. In low measurement regime, proposed perturbation approach gives lower NMSE-H results with increasing number of snapshots, while at higher number of snapshots the achieved channel estimation performance at a single snapshot is consistent for increased number of snapshots as well.

\subsection{Performance of Different Covariance Estimation Algorithms}

\begin{figure}[!t]
\centerline{\includegraphics[scale=0.65]{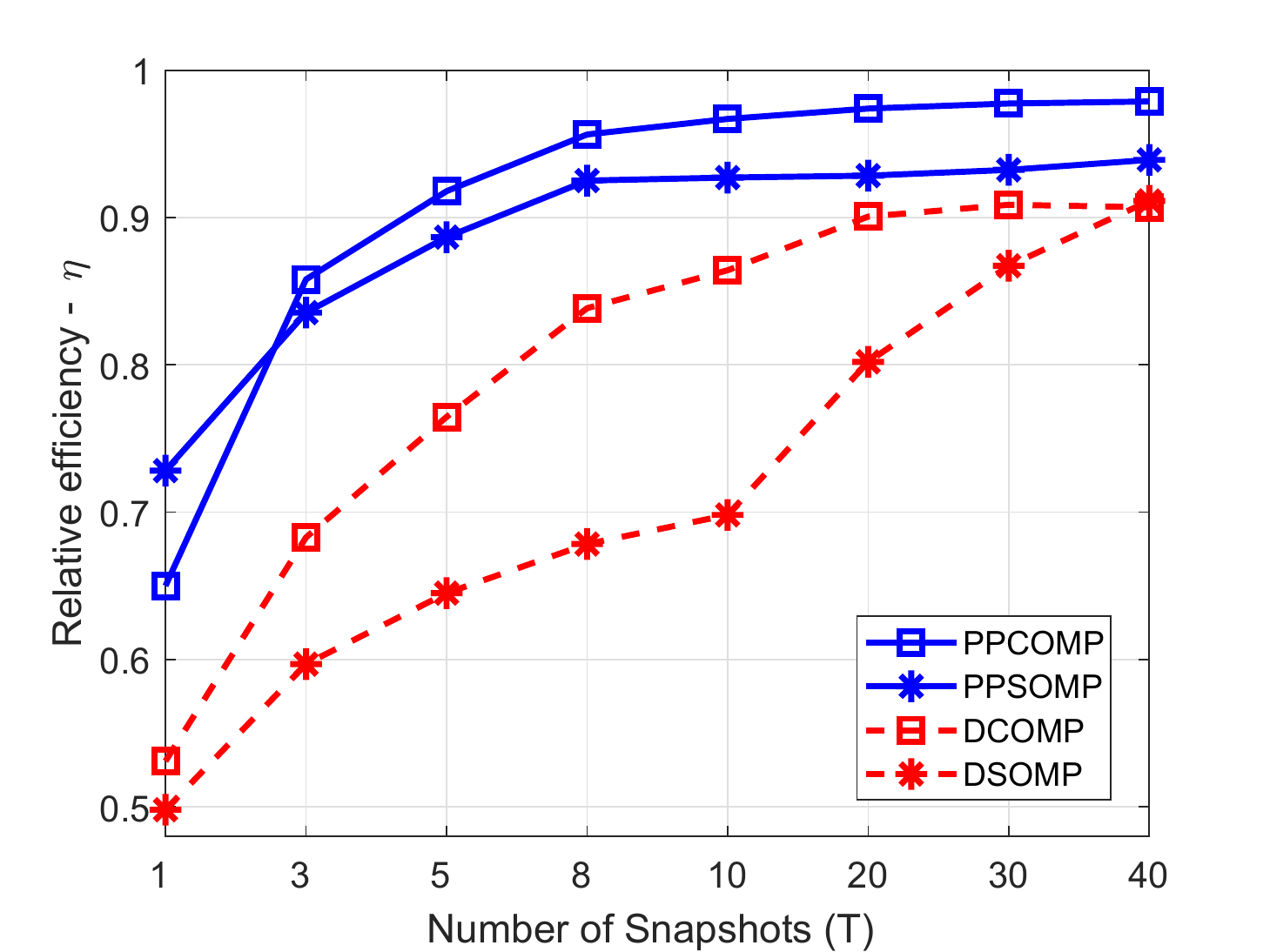}}\vspace{-2mm}
\caption{Comparison of relative efficiency between different methods versus the number of snapshots ($M_\text{RF}N_\text{RF}$ = 30, SNR = 10 dB, and uniform sampling of $\cos(\theta)$ domain).}
\label{Fig:Eta_Algo_Comparison}
\end{figure}

\begin{figure}[!t]
\centerline{\includegraphics[scale=0.65]{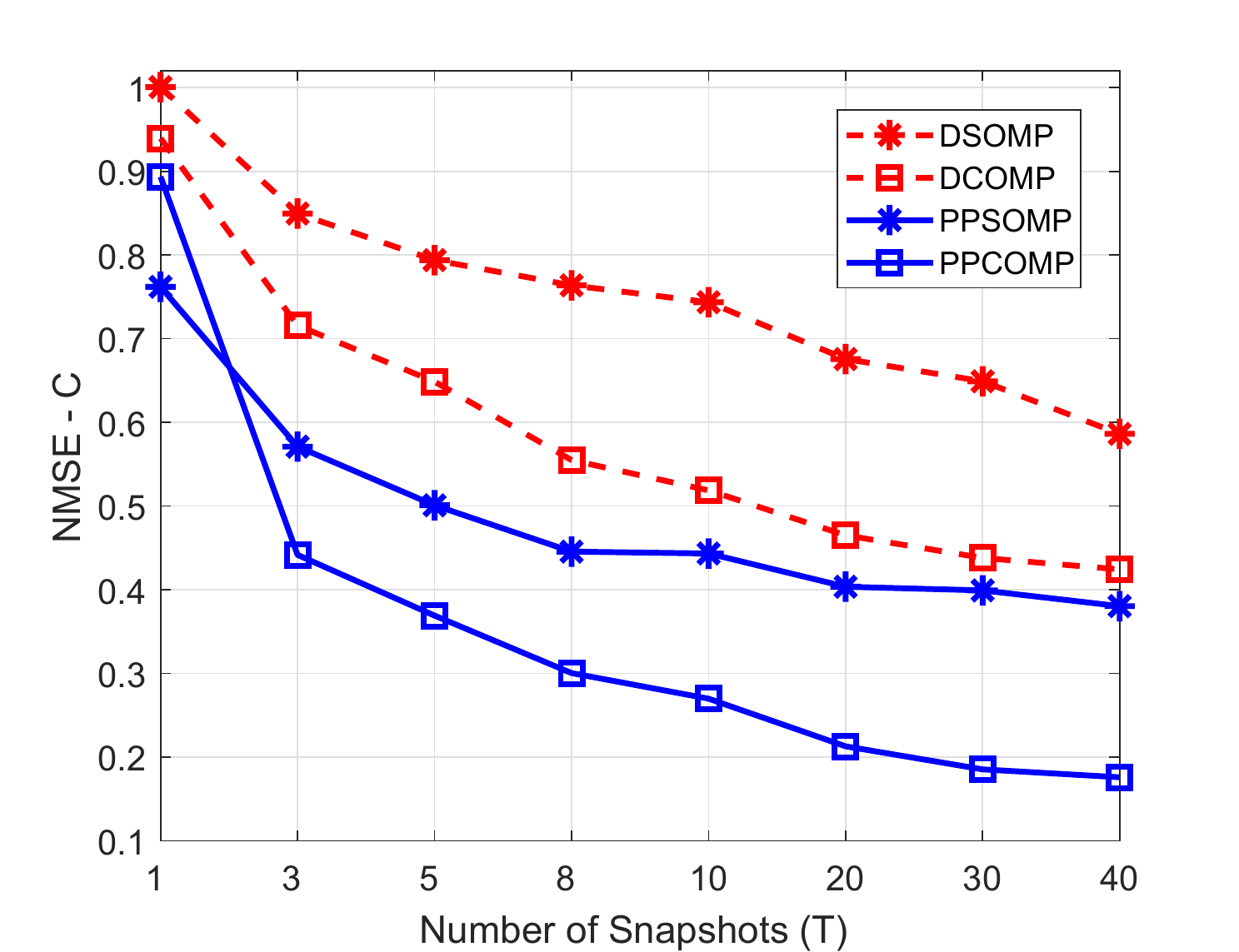}}\vspace{-2mm}
\caption{Comparison of NMSE-C between different methods versus the number of snapshots ($M_\text{RF}N_\text{RF}$ = 30, SNR = 10 dB, and uniform sampling of $\cos(\theta)$ domain).}
\label{Fig:Err_C_Algo_Comparison}
\end{figure}

Fig. \ref{Fig:Eta_Algo_Comparison} compares the performance of different covariance estimation algorithms in terms of relative efficiency $\eta$ with $M_\text{RF}N_\text{RF}$ = 30 and SNR = 10 dB. It can be seen that the parameter perturbed algorithms PPCOMP and PPSOMP outperforms DCOMP and DSOMP, respectively. The performance improvement of PPCOMP and PPSOMP is due to the fact that it is better equipped to capture the off-grid by means of controlled perturbed mechanism, whereas the DCOMP and DSOMP fails to do so. It is also observed that parameter perturbed algorithms reach their peak performance at a smaller number of snapshots, which reduces the estimation time for fast changing enviroments in mmWave applications. On the other hand, the counterpart algorithms require relatively more snapshots to reach its peak performance which is lower than the perturbed versions. Among the perturbed algorithms, the PPCOMP performs relatively better than the PPSOMP as it is more robust to variations to the channel dynamics as compared to estimation of the instantaneous channel coefficients \cite{Beam_Alignment_Song_Caire}. 

Fig. \ref{Fig:Err_C_Algo_Comparison} shows the performance of different covariance algorithms in terms of NMSE-C for the same simulation parameters. A similar trend, where the perturbed algorithms PPCOMP and PPSOMP outperforms other algorithms is also observed for the NMSE-C metric. Here onwards, we restrict our discussion to the PPCOMP and its counterpart DCOMP algorithm for evaluating the covariance algorithm performance since they outperform the SOMP based techniques. 

\subsection{Impact of Employed Sampling Scheme }
\begin{figure}[t!]
\centerline{\includegraphics[scale=0.5]{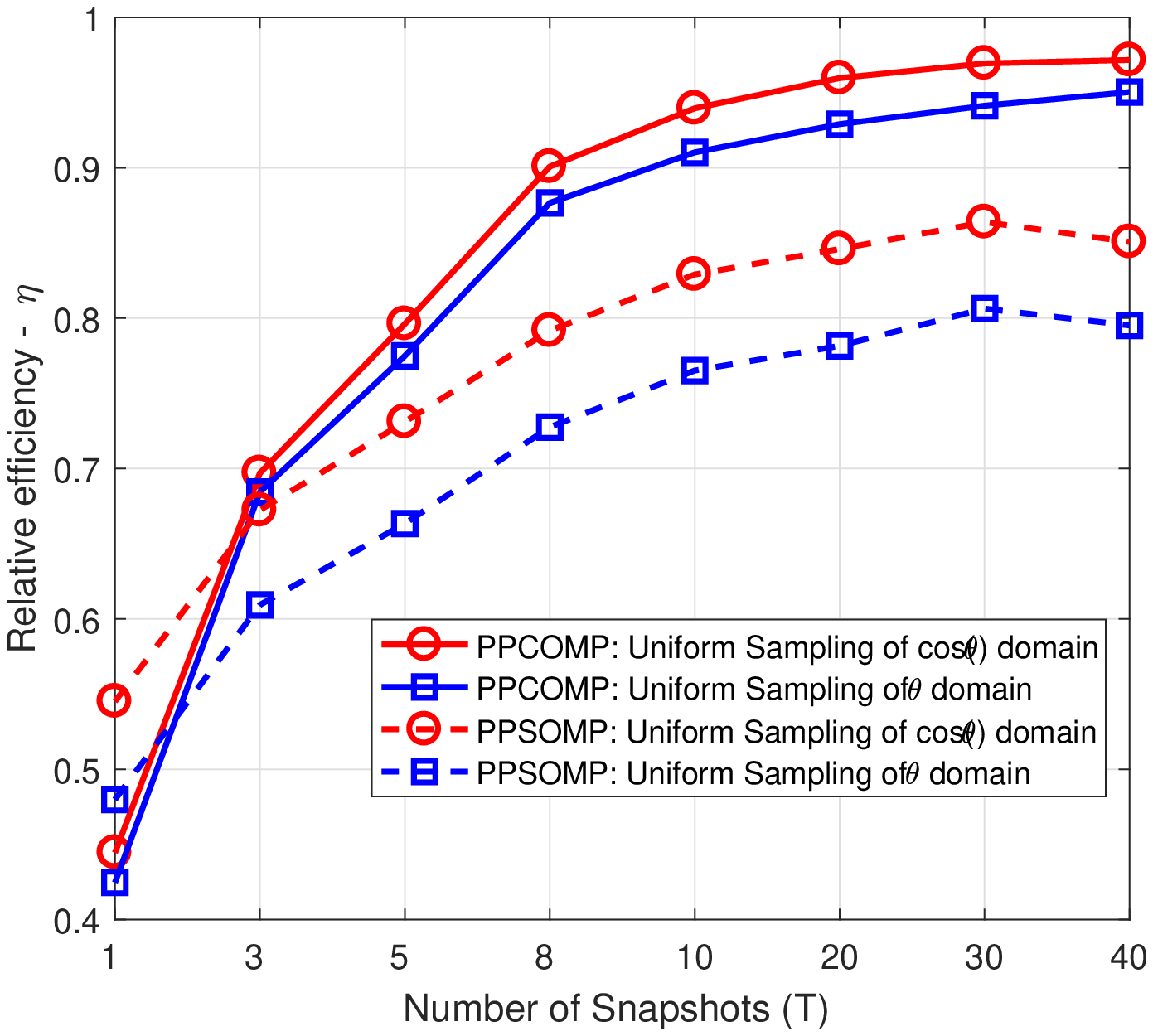} \vspace{-2mm} \includegraphics[scale=0.6]{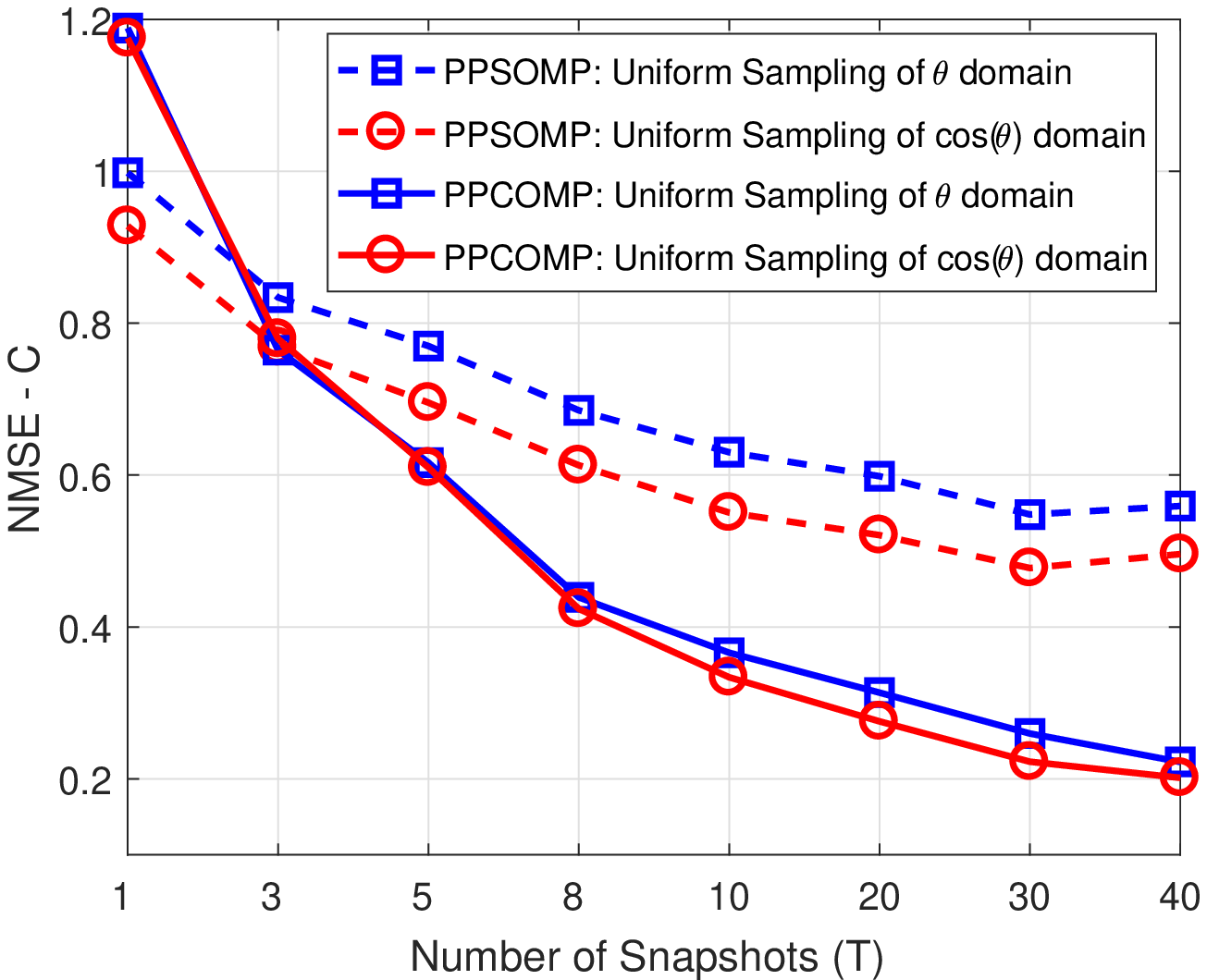}}
\caption{Relative efficiency (Left) and NMSE-C (Right) performance based on the sampling scheme employed ($M_{RF}N_{RF}$ = 30 and SNR = 10 dB).}
\label{Fig:Sampling_Eta}
\end{figure}

As previously indicated, the sampling scheme employed significantly influences the performance of the sparse recovery techniques. Fig. \ref{Fig:Sampling_Eta} illustrates the impact of two sampling schemes discussed in Section \ref{SubSub:Discretization} on the performance of $\eta$ and NMSE-C. The performance is evaluated only for the perturbed algorithms which are shown to be superior in terms of performance compared to the non-perturbed algorithms as established in the previous subsection. From Fig.~\ref{Fig:Sampling_Eta}, it can be seen that employing \textit{uniform sampling of the virtual domain} rather than the classical approach of \textit{uniform sampling of the physical domain} improves the performance of the covariance estimation. This can be attributed to the fact that the former scheme reduces the mutual coherence between the discretized points which in turn helps to find the better initial grid points for the class of perturbed algorithms. Other presented results employs uniform sampling of $\cos(\theta)$ domain due to its increased performance.
 
\subsection{Effect of Discretization Level}

\begin{figure}[!t]
\centerline{\includegraphics[scale=0.7]{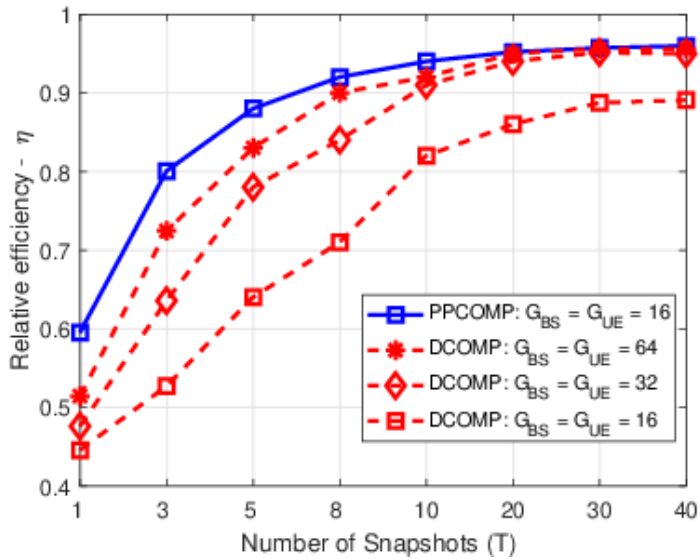}}\vspace{-2mm}
\caption{Impact of the grid size employed on the relative efficiency ($M_{RF}N_{RF}$ = 30, SNR = 10 dB, and uniform sampling of $\cos(\theta)$ domain).}
\label{Fig:Grid_Effect_Eta}
\end{figure}

In this subsection, we investigate the effect of number of grid points on the algorithms performance. For this purpose, we use the PPCOMP algorithm with $G_{BS} = G_{UE} = 16$ as the benchmark case and evaluate the performance of DCOMP algorithm with increasing number of grid points. The number of measurements was fixed to $M_{RF}N_{RF}$ = 30, SNR = 10 dB, and uniform sampling of $\cos(\theta)$ domain. It can be observed from Fig. \ref{Fig:Grid_Effect_Eta} that increasing the number of grid points (the level of discretization ) can increase the performance of the DCOMP algorithm as it reduces the error caused due to the basis mismatch. Even though increasing the number of grid points has a positive effect, it also has negative effects. As noted before, it increases the mutual correlation  of the dictionary matrix and also leads to the undesirable increase in the computational complexity. To conclude, rather than using DSOMP/DCOMP over a larger and denser dictionary, it is advisable to use PPPOMP over a much smaller size dictionary \cite{teke2013perturbed}.

\subsection{Effect of Different SNR Levels}

\begin{figure}[!t]
\centerline{\includegraphics[scale=0.65]{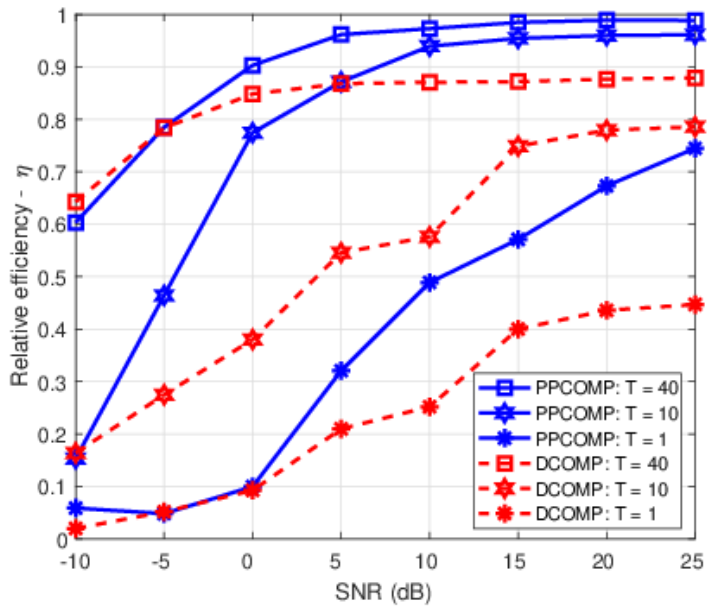}}\vspace{-2mm}
\caption{Comparison of relative efficiency of PPCOMP and DCOMP at different SNR level ($M_\text{RF}N_\text{RF}$ = 30, and uniform sampling of $\cos(\theta)$ domain).}
\label{Fig:SNR_Eta}
\end{figure}

To investigate the effect of different SNR levels, we vary the SNR range  from -10 dB to 25 dB and evaluate the performance of the DCOMP and PPCOMP algorithms at 3 different snapshots level ($T$ = 1, 10, and 40). For this simulation, the number of measurements are fixed to $M_\text{RF}N_\text{RF}$ = 30 with uniform sampling of $\cos(\theta)$ domain. As seen in Fig. \ref{Fig:SNR_Eta}, at lower SNR regime (in the range of -10 to 0 dB), the performance of PPCOMP and DCOMP are almost comparable as both algorithms have lower efficiency levels due to not able to find the correct support. However, beyond medium SNR levels (beyond 5 dB), the PPCOMP exhibits increased efficiency compared to DCOMP at the same snapshot level. The PPCOMP even performs at higher efficiency with 10 snapshot compared to DCOMP with 40 snapshots for 5dB or higher SNR levels. Even PPCOMP with 1 snapshot performs nearly as DCOMP with 10 snapshots. In summary, for high enough SNRs, the proposed technique allows similar performance with lower number of snapshots. 

\subsection{Dependence on the Number of Measurements}

\begin{figure}[!t]
\centerline{\includegraphics[scale=0.75]{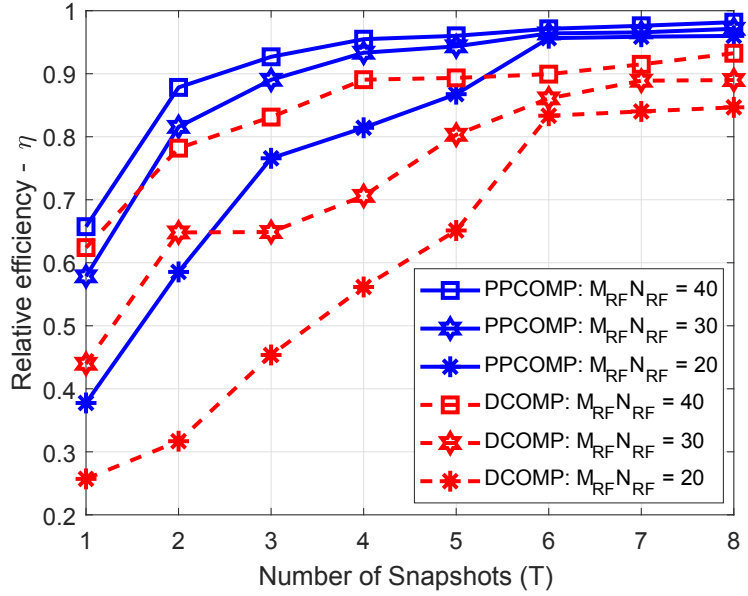}}\vspace{-2mm}
\caption{Relative efficiency performance with different measurements (SNR $=10$~dB and uniform sampling of $\cos(\theta)$ domain).}
\label{Fig:Measurements}
\end{figure}

Like the number snapshots, the number of measurements (RF chains) also significantly influences the performance of covariance estimation algorithms. Fig. \ref{Fig:Measurements} investigates the impact of measurements ($M_\text{RF}N_\text{RF}$ = 20, 30, and 40) on the relative efficiency metric as a function of the number of snapshots. Fig. \ref{Fig:Measurements} suggests a trade off between measurements and snapshots. The general trend is that with smaller number of measurements, the algorithms require more snapshots to reach their peak performance. While for even increased number of snapshots DCOMP efficiency converges to different levels, PPCOMP is able to provide a higher efficiency levels for all tested measurement number cases with increasing number of snapshots.  thereby a lesser number of measurements and snapshots are required for covariance estimation using PPCOMP. 

\subsection{Dependence on the Number of Antennas}

\begin{figure}[!t]
\centerline{\includegraphics[scale=0.65]{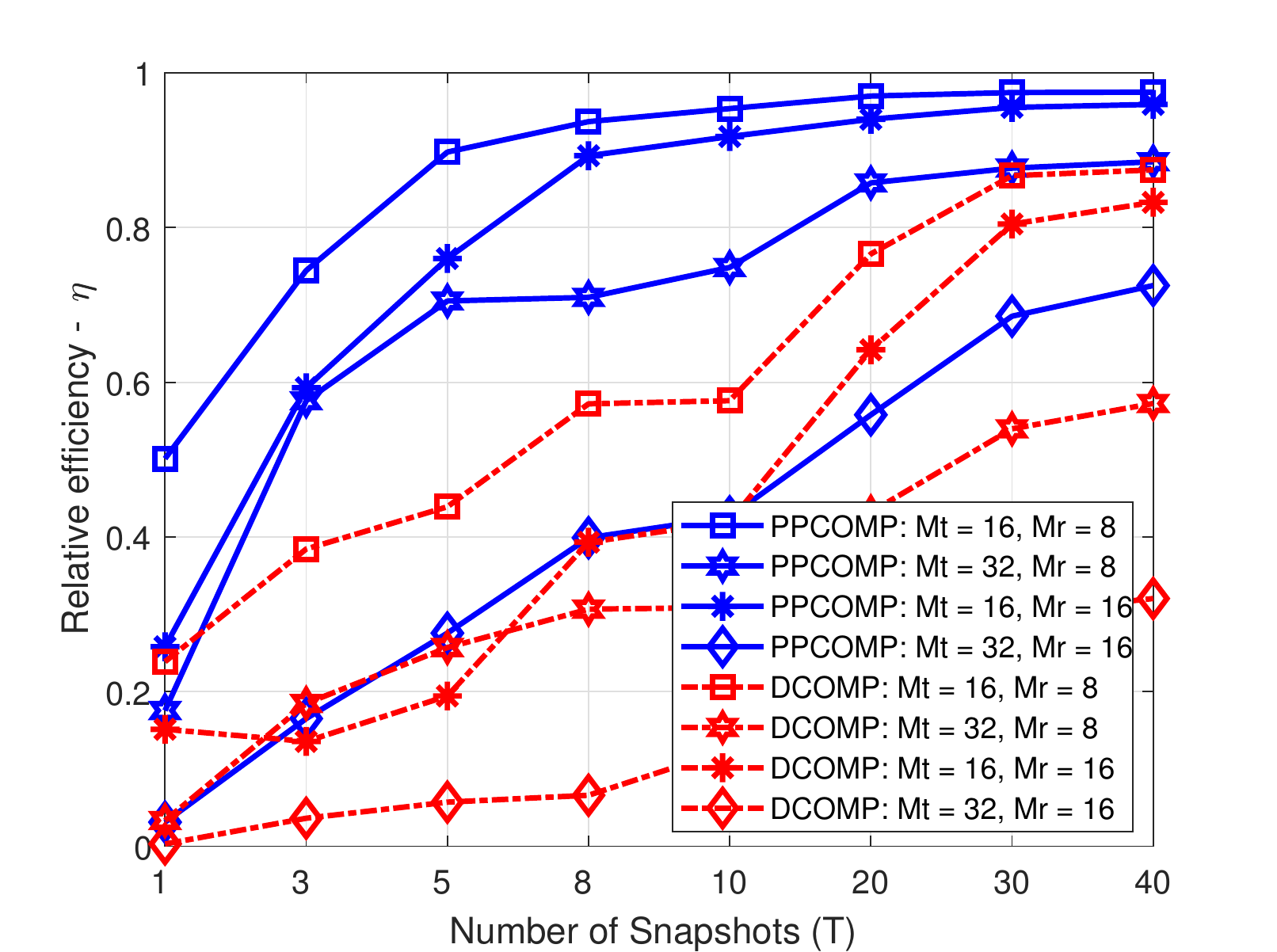}}
\caption{Relative efficiency dependence on the number of antennas for the PPCOMP and COMP algorithms with the number of measurement (RF chains) fixed to $M_\text{RF}N_\text{RF} = 30$, SNR = 10 dB, and uniform sampling of $\cos(\theta)$ domain.}
\label{Fig:Number_of_Antennas}
\end{figure}

Fig. \ref{Fig:Number_of_Antennas} illustrates the effect of varying the number of antennas at the BS and UE with the RF chains fixed in the system. As evident, the relative efficiency metric degrades with increase in the number of antennas at the BS and UE. This degradation is severe for the DCOMP algorithm. On the other hand, the PPCOMP still maintains the superiority with a significant difference in the performance due to the controlled perturbation scheme which evades the off-grid effects and improves the overall performance significantly requiring lesser snapshots and measurements.

\section{Conclusion}\label{sec:conclusion}
In this paper, we study the channel estimation and covariance estimation problems for MIMO mmWave network setup considering the off-grid effects. We propose the PPSOMP and PPCOMP algorithms for the explicit channel estimation and covariance estimation, respectively. The proposed algorithms evade the issue arising from the basis mismatch problems by operating on the continuum AoA-AoD space using the mechanism of the controlled perturbation in conjunction with a modified SOMP framework. The modified SOMP framework helps to preserve the low computational complexity which is inherent for a greedy solver. On the other hand, the controlled perturbation mechanism jointly solves for the off-grid parameters and weights. Simulation results demonstrate the superiority of our proposed methods, and  outperforms the existing techniques both in terms of the relative efficiency metric and reconstruction error. 

\bibliographystyle{IEEEtran}
\bibliography{IEEEabrv,papers}

\end{document}